\documentclass[a4paper]{article}
\usepackage{amsmath}
\usepackage{amssymb}
\usepackage[pdfborder={0 0 0}]{hyperref}
\usepackage[all]{hypcap}
\usepackage{lineno}
\usepackage{array}
\usepackage{calc}
\usepackage{longtable}
\usepackage{multirow}
\usepackage{graphicx}
\usepackage{xspace}
\usepackage{tikz}
\usetikzlibrary{arrows.meta,decorations.markings,angles,quotes}
\usepackage{cite}
\usepackage{color}
\usepackage[format=hang,labelfont=bf]{caption}
\usepackage{sectsty}
\usepackage{booktabs}

\allsectionsfont{\sffamily}
\subsubsectionfont{\mdseries\itshape\large}
\makeatletter
\let\spreprint\@empty
\newcommand{\preprint}[1]{\gdef\spreprint{#1}}
\let\sinstitute\@empty
\newcommand{\institute}[1]{\gdef\sinstitute{#1}}
\def\@maketitle{%
  \newpage
  \null
  \ifx\spreprint\@empty\else
    \makebox[0pt][l]{%
      \raisebox{2\baselineskip}[0pt][0pt]{%
        \hspace{\textwidth}%
        \llap{\normalsize\shortstack[r]{\spreprint}}}}%
  \fi
  \vskip 2em%
  \begin{center}%
    \let\footnote\thanks
    {\LARGE \@title \par}%
    \vskip 1.5em%
    {\large
      \lineskip .5em%
      \begin{tabular}[t]{c}%
        \@author
      \end{tabular}\par}%
    \ifx\sinstitute\@empty\else
      \vskip 0.9em%
      {\normalsize \sinstitute \par}%
    \fi
    \vskip 1em%
    {\large \@date}%
  \end{center}%
  \par
  \vskip 1.5em}
\makeatother
\renewenvironment{abstract}{\begin{center}
  {\large\sffamily\bfseries Abstract: }
  \begin{minipage}[t]{0.75\textwidth}
}{\end{minipage}\end{center}\vskip 10ex}

\newcommand{\Herwig}{H\protect\scalebox{0.8}{ERWIG}\xspace}

\newcommand{\Sherpa}{S\protect\scalebox{0.8}{HERPA}\xspace}

\newcommand{\Comix}{C\protect\scalebox{0.8}{OMIX}\xspace}

\newcommand{\Amegic}{A\protect\scalebox{0.8}{MEGIC++}\xspace}
\newcommand{\Rivet}{R\protect\scalebox{0.8}{IVET}\xspace}

\newcommand{\Pythia}{P\protect\scalebox{0.8}{YTHIA}\xspace}

\newcommand{\Superchic}{S\protect\scalebox{0.8}{uperCHIC}\xspace}
\newcommand{\Starlight}{S\protect\scalebox{0.8}{TARlight}\xspace}
\newcommand{\UPCgen}{U\protect\scalebox{0.8}{PC}gen\xspace}

\newcommand{\MadgraphaMCatNLO}{M\protect\scalebox{0.8}{ad}G\protect\scalebox{0.8}{raph5\_a}MC\scalebox{0.8}{at}NLO\xspace}
\long\def\symbolfootnote[#1]#2{\begingroup%
\def\thefootnote{\fnsymbol{footnote}}\footnote[#1]{#2}\endgroup}

\newcommand{\beq}{\begin{equation}}
\newcommand{\eeq}{\end{equation}}
\newcommand{\bal}{\begin{align}}
\newcommand{\eal}{\end{align}}
\newcommand{\done}{{\rm d}}

\newcommand{\dthree}{{\rm d}^3}

\newcommand{\GeV}{\mathrm{GeV}}

\newcommand\EIC{E\scalebox{0.8}{IC}\xspace}

\newcommand\LHC{L\protect\scalebox{0.8}{HC}\xspace}
\newcommand\ATLAS{\atlas}
\newcommand\atlas{A\protect\scalebox{0.8}{TLAS}\xspace}

\usepackage{cleveref}

\title{Automated higher-order predictions for ultra-peripheral collisions with full impact-parameter dependence}
\author{Frank~Krauss$^{a}$, Peter~Meinzinger$^{b}$}
\date{}
\preprint{ZU-TH 26/26 \\ IPPP/26/57}
\institute{%
  $^{a}$\,Institute for Particle Physics Phenomenology, Durham University,
  Durham DH1 3LE, United Kingdom\\[2pt]
  $^{b}$\,Physik-Institut, Universit\"at Z\"urich,
  Winterthurerstrasse 190, 8057 Z\"urich, Switzerland%
}

\begin{document}

\maketitle
\begin{abstract}
  We present an automated framework within the \Sherpa event generator for the simulation of photon-induced
  processes in high-energy collisions of protons and nuclei.
  The calculation builds upon the equivalent-photon approximation and incorporates
  different form factor parametrisations, taking into account the full
  dependence on the finite size of emitters and their impact parameter.
  Higher-order effects are taken into account automatically by
  either YFS-resummation of soft photon emissions to all orders
  or Next-to-Leading Order in electroweak theory.
  We validate our simulation framework using data from two representative
  \ATLAS measurements of exclusive $\mu$-pair production in $pp$ and Pb+Pb collisions.
  Exemplary predictions for exclusive $W$-pair production in $pp$ collisions and
   $\tau$-pair production in both $pp$ and Pb+Pb collisions are presented.
\end{abstract}

\section{Introduction}
Ultra-peripheral high-energy collisions (UPCs) of protons or nuclei have received renewed attention at the Large Hadron Collider (\LHC).
They are characterized by unique experimental signatures, where the two incident beam particles effectively ``miss'' each other and stay intact,
and where additional final state particles are produced through mostly electromagnetic interactions.
UPCs turn the \LHC effectively into a photon collider and offer a wide range of new opportunities:
The observation of rare quantum effects such as light-by-light scattering and searches for axion and axion-like particles~\cite{CMS:2018erd,ATLAS:2020hii} or
measurements of (and constraints on) the anomalous magnetic moment of the $\tau$ lepton~\cite{ATLAS:2022ryk,CMS:2024qjo} serve as illustrative examples.
The description of photon-induced processes in UPCs usually employs Fermi's method of virtual quanta~\cite{Fermi:1924tc},
with Weizsaecker-Williams spectra of the equivalent quasi-real photons~\cite{vonWeizsacker:1934nji,Williams:1934ad} as a starting point.
Their semi-classical method has been more rigorously derived in Budnev et al.~\cite{Budnev:1975poe},
including polarization and off-shell effects in the equivalent photon approximation (EPA).
Reviews, for instance~\cite{Bertulani:1987tz,Krauss:1997vr,Bertulani:2005ru}, summarize its early applications.

In recent years a range of new specialized simulation tools have addressed UPCs.
The \Starlight code~\cite{Klein:2016yzr} models the production of lepton pairs, selected single mesons, and meson pairs in photon-photon and
photo-nuclear reactions and includes various models for the nuclear survival or breakup as a result of the interaction.
Similar to \Starlight, \UPCgen~\cite{Burmasov:2021phy} simulates the production of lepton pairs and other final states in two-photon processes.
\Superchic~\cite{Harland-Lang:2020veo} traditionally focused on the simulation of exclusive heavy quarkonia production in $pp$ collisions
at the \LHC with a more theory-based precise description of survival factors, see e.g.~\cite{Harland-Lang:2021ysd}.
Since its inception, this Monte Carlo tool further extended its scope and now includes an increasing array of possible
final states in the interactions in various elastic, inelastic, and diffractive production modes.
This is exemplified by studies of $\tau$ anomalous magnetic moment~\cite{Harland-Lang:2024zpn} and $W$-pair production~\cite{Bailey:2022wqy} measurements at the \LHC,
which focus in particular on the observable impact of the survival or dissociation of the protons.
All of these specialized tools use multi-purpose event generators as
``afterburners'' -- mainly \Pythia~\cite{Sjostrand:2006za,Bierlich:2022pfr} -- for additional physics effects in the final states,
such as parton fragmentation and hadronization, hadron decays or QED final state radiation (FSR).

Parallel to these tool developments, higher-order corrections for a range of photon-induced processes
have been calculated, for example next-to-leading order (NLO) QCD corrections to inclusive dijet
photo-production~\cite{Guzey:2018dlm},  NLO QED corrections to $\mu^-\mu^+$ and $\tau^-\tau^+$ production~\cite{Shao:2024dmk},
or, more recently, the full NLO electroweak correction to $\tau^+\tau^-$ production~\cite{Jiang:2024dhf,Dittmaier:2025ikh}
in UPCs in heavy-ion collisions at the \LHC.

Note that such NLO corrections are also available in automated tools
such as \MadgraphaMCatNLO~\cite{Shao:2024dmk}, which,
combined with EPA spectra and through their link to multi-purpose event generators
such as \Pythia or \Herwig~\cite{Bellm:2025pcw}, provide precise full particle-level simulations
for an increasing range of processes.
In this paper we report on similar efforts within the \Sherpa framework.

This paper is organized as follows:
In Sec.~\ref{Sec:EPA} we introduce the relevant kinematics and summarize the EPA formalism
underpinning our simulation in \Sherpa and list the available form factor parametrizations.
There, we also discuss the impact-parameter dependence of the spectra and various ideas and approximations to estimate the effect of additional
interactions of the projectiles which would lead to their break-up and thereby obscure the experimental signature.
In Sec.~\ref{Sec:Validation} we compare the results of the \Sherpa simulation with existing \LHC data
for the production of $\mu^-\mu^+$ pairs in proton-proton and lead-lead collisions.
In this section we focus on the impact of different form factor parametrizations on observable muon-pair mass spectra,
and we quantify the effect of different ways to estimate the proton and lead survival probabilities.
We also include higher-order corrections in two different ways, namely through a Next-to-Leading Order
electroweak calculation and through the resummation of soft photons to all orders in YFS formalism~\cite{Yennie:1961ad}.
In Sec.~\ref{Sec:Predictions} we show predictions for two exemplary processes within reach of current experimental capabilities and efforts,
namely exclusive $W$- and $\tau$-pair production.
We conclude in Sec.~\ref{Sec:Conclusion} with a summary of the key findings and results of this paper and an outlook to further studies. The study of photon-induced jet final states at the \LHC will be discussed in a separate publication~\cite{photonuclear-jets}.

\section{Photon spectra}
\label{Sec:EPA}
\subsection{Kinematics}

The photon virtuality can be parametrised in the following way.
Consider a beam particle with mass $m$, incoming energy $E$ moving in the $z$-direction,
outgoing energy $E^\prime$, and transverse momentum $k_\perp$ generated in the emission of the virtual photon.
The virtuality $Q^2 = - q^2$ of the photon with momentum $q$ is given by

\begin{align}
    Q^2 &= -2 m^2 + 2 \left( E E^\prime - \sqrt{E^2 - m^2} \sqrt{E^{\prime 2} - k_\perp^2 - m^2} \right) \\
    &\approx \frac{m^2 k_\perp^2 (x-2) x}{2 E^2 (x-1)^3}+\frac{m^2 x^2+k_\perp^2}{1-x} + \mathcal{O} \left(\frac{m^4}{E^4},\ \frac{m^4}{E^{\prime 4}} ,\ \frac{k_\perp^4}{E^{\prime 4}}\right)\\
    &\approx \frac{m^2 x^2+k_\perp^2}{1-x} + \mathcal{O}\left(\frac{1}{E^2}\right)
    \label{eq:q2-approx}
\end{align}

where we used $E^\prime = (1-x) E$, expanded around $m/E^{(\prime)} \to 0$ and $k_\perp/E^\prime \to 0$ in the first step,
and took the high-energy limit ($E \to \infty$) in the second step.
Throughout this work, we use only the final approximation.

We denote by $x$ the fractional energy carried by the photon with respect to the emitting particle,
\begin{equation}
    x \equiv \frac{E_\gamma}{E} = 1 - \frac{E^\prime}{E} \, ,
\end{equation}
where $E_\gamma$ is the photon energy in the laboratory frame.
The kinematic limits on $Q^2$ are given by
\begin{align}
    Q^2_\mathrm{min}(x) &= \frac{m^2 x^2}{1-x} \\
    Q^2_\mathrm{max}(x) &= \frac{k_{\perp,\mathrm{max}}^2 + m^2 x^2}{1-x}
\end{align}
where $k_{\perp,\mathrm{max}}$ corresponds to the maximal scattering angle of the beam particle.

\subsection{General formalism}

The photon flux can be derived from the spin density matrix of the photon~\cite{Budnev:1975poe}
after applying the approximation $Q^2 \ll W^2$ (where $W$ is the invariant mass of the produced system)
and averaging over the azimuthal angle $\varphi$.
This yields the unpolarized differential photon flux

\begin{align}
    \frac{\dthree n(x,\,Q^2,\,\varphi)}{\done x\,\done Q^2\,\done \varphi} &=
    \frac{\alpha}{2\pi^2}\,\frac{1}{xQ^2}\,\left[
    \frac{x^2}{2}\,C(Q^2)+
    \left(1-x\right)\left(1-\frac{Q^2_{\rm min}}{Q^2}\right)\,D(Q^2)\right] \nonumber \\
    &= \frac{\alpha}{2\pi^2}\,\frac{1}{xQ^2}\,\left[\frac{x^2}{2}\,C(Q^2)+\frac{k_\perp^2}{Q^2}\,D(Q^2) \right]
    \label{eq:general-flux}
\end{align}

where $\alpha$ is the fine-structure constant.
The coefficients $C(Q^2)$ and $D(Q^2)$ encode the magnetic and
electric form factor contributions, respectively, and depend on the particle type.

For heavy nuclei, the magnetic form factor contribution can be neglected, setting $C(Q^2) = 0$.
This approximation is justified because magnetic contributions are suppressed by $\mathcal{O}(Q^2/m^2)$
compared to electric contributions and are negligible in the kinematic regime of interest.
For the electric form factor, we write
\begin{equation}
    D(Q^2) = Z^2 \left|F(Q^2)\right|^2
\end{equation}
where $Z$ is the charge number of the emitting particle and
$F(Q^2)$ is the electromagnetic form factor normalized such that $F(0) = 1$.

\begin{figure}[htbp]
\centering
\begin{tikzpicture}[>=Stealth]
    \coordinate (A1) at (-2.5,0);
    \coordinate (A2) at (2.5,0);
    \coordinate (collision) at (0.5,1.5);

    \fill[gray!30] (A1) circle (1.1cm);
    \draw[thick] (A1) circle (1.1cm);
    \fill[gray!30] (A2) circle (1.1cm);
    \draw[thick] (A2) circle (1.1cm);

    \node[below=5pt] at (A1) {$A_1$};
    \node[below=5pt] at (A2) {$A_2$};

    \draw[->,black] (A1) -- (collision) node[midway,above] {$b_{\gamma,1}$};
    \draw[->,black] (A2) -- (collision) node[midway,above] {$b_{\gamma,2}$};

    \fill (collision) circle (2pt);
    \node[above=1pt] at (collision) {$\gamma \gamma \to X$};

    \draw pic [draw, angle radius=0.4cm, angle eccentricity=1.6, "$\phi$"] {angle = A1--collision--A2};
\end{tikzpicture}
\caption{Setup for photon-photon collisions in the impact parameter space.
The impact parameters $b_{\gamma,1}$ and $b_{\gamma,2}$ represent the transverse distances
between the photons and their respective emitting nuclei $A_1$ and $A_2$.}
\label{fig:geometric-setup}
\end{figure}
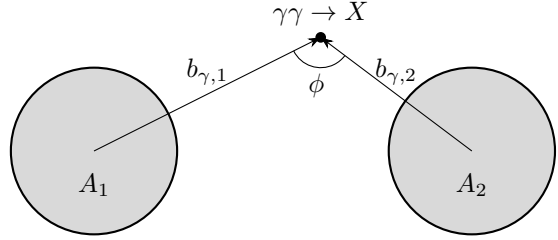

The photon flux can be expressed as a function of the
impact parameter $b_\gamma$ (the transverse distance between photon and the emitting particle)
by Fourier transforming Eq.~\ref{eq:general-flux} with respect to
$k_\perp$~\cite{Vidovic:1993cf,Vidovic:1992ik}, which gives

\begin{equation}
    n (x, b_\gamma) = \frac{Z^2 \alpha}{\pi^2 x} \left| \int_0^\infty \done k_\perp \,
    \frac{k_\perp^2}{Q^2} \, F (Q^2) \, J_1 (b_\gamma k_\perp) \right|^2
    \label{eq:fourier-transform}
\end{equation}

Here $J_1$ is the Bessel function of the first kind, and
$Q^2$ implicitly depends on both $k_\perp$ and $x$ through Eq.~\ref{eq:q2-approx}.

For non-trivial form factors the integral will not yield a closed analytic expression and
Eq.~\ref{eq:fourier-transform} must be evaluated numerically.
The oscillatory nature of the $J_1$ function renders this non-trivial.
We employ a custom implementation of the adaptive integration method of Refs.~\cite{sidi1988user,lucas1995evaluating},
which we have validated against known analytic cases.

\subsection{Form factor parametrizations}

We implement several form factor parametrizations, which we describe below.
For each form factor, we compute the photon flux via Eq.~\ref{eq:fourier-transform} unless an analytic expression exists.
Table~\ref{tab:form-factors} summarizes the form factors and their typical applications.

\begin{table}[htbp]
\centering
\begin{tabular}{llll}
\toprule
Form factor & Application & Evaluation & $b$-dependent \\
\midrule
point-like & Proton, large-$b$ fallback & Analytic & \checkmark \\
integrated point-like & Proton & Analytic & - \\
proton & Protons & Analytic & - \\
proton, approx. & Protons & Analytic & - \\
dipole & Protons, light ions & Numerical & \checkmark \\
approx. dipole & Protons, light ions & Numerical & \checkmark \\
Gaussian & Phenomenological & Numerical & \checkmark \\
hom.-charged sphere & Phenomenological & Numerical & \checkmark \\
Woods-Saxon & Heavy ions & Numerical & \checkmark \\
Woods-Saxon, approx. & Heavy ions & Numerical & \checkmark \\
\bottomrule
\end{tabular}
\caption{Summary of implemented form factors. Numerical evaluation refers to the integration of Eq.~\ref{eq:fourier-transform}.}
\label{tab:form-factors}
\end{table}

\paragraph{Point-like approximation}

Using the approximation of a point-like source for nuclei, the flux is given by

\begin{equation}
    n_\mathrm{PL} (x, b_\gamma) = \frac{2 Z^2 \alpha}{\pi} x m_\mathrm{N}^2 \left[ K_1^2(\chi) + \frac{1}{\gamma} K_0^2(\chi) \right]
    \label{eq:point-like-flux}
\end{equation}

where $\chi = x m_\mathrm{N} b_\gamma$, $\gamma$ is the Lorentz factor,
$K_{0,1}$ are modified Bessel functions of the second kind and $m_\mathrm{N} = m_A / A$ is the nucleon mass.
The first term represents transversely polarized photons,
while the second, suppressed term accounts for longitudinally polarized photons.
At asymptotically high energies, the longitudinal contribution becomes negligible.

For this flux, the impact parameter $b_\gamma$ can be integrated out analytically, yielding:

\begin{equation}
    n_\mathrm{PL,int} (x) = \frac{2 Z^2 \alpha}{\pi x} \left[ \chi \, K_0(\chi) K_1(\chi)
    - \frac{\chi^2}{2} \left( K_1^2(\chi) - K_0^2(\chi) \right) \right]
    \label{eq:ion-approx-integrated}
\end{equation}

where now $\chi = x m_\mathrm{N} b_{\gamma,\mathrm{min}}$ with $b_{\gamma,\mathrm{min}} = \max(R, \tilde{b}_\mathrm{min} R)$.

\paragraph{Budnev proton parametrisation}

For the proton, we use the form factor from Ref.~\cite{Budnev:1975poe} (see also below), with $Q_0^2 = 0.71$ GeV$^2$.
The photon flux can be calculated via Eq.~\ref{eq:general-flux},
and when integrated over the virtualities one yields~\cite{Budnev:1975poe}:

\begin{align}
    n_\mathrm{Budnev} (x) &= \frac{\alpha(1-x)}{\pi x}
    \left[ \varphi\left(\frac{Q^2_\mathrm{max}}{Q_0^2}, x\right) - \varphi\left(\frac{Q^2_\mathrm{min}}{Q_0^2}, x\right) \right]
    \label{eq:proton-flux} \ \mathrm{with} \\
    \varphi(z, x) &= (1 + a y) \left[ -\log\left(1 + \frac{1}{z}\right) + \frac{1}{1+z}
    + \frac{1}{2(1+z)^2} + \frac{1}{3(1+z)^3} \right] \nonumber \\
    &\quad + \frac{(1-b) y}{4z(1+z)^3} \nonumber \\
    &\quad + c \left(1 + \frac{y}{4}\right) \left[ \log\left(\frac{1+z-b}{1+z}\right) + \frac{b}{1+z}
    + \frac{b^2}{2(1+z)^2} + \frac{b^3}{3(1+z)^3} \right]
\end{align}

where we use the proton's magnetic moment $\mu_p = 2.79$~\cite{ParticleDataGroup:2024cfk},
and use $y = \frac{x^2}{1-x}$, $a = \frac{1 + \mu_p^2}{4} + \frac{4 m^2}{Q_0^2}$,
$b = 1 - \frac{4 m^2}{Q_0^2}$ and $c = \frac{\mu_p^2 - 1}{b^4}$.

An approximate version, valid for the region $Q^2 \to 0\ \GeV^2$, sets $F(Q^2) \to 1$ and yields

\begin{equation}
    n_\mathrm{Budnev, approx} (x) = \frac{\alpha}{\pi x}
    \left[ \left(1 - x + \frac{\mu_p^2 x^2}{2}\right) \log\left(\frac{Q^2_\mathrm{max}}{Q^2_\mathrm{min}}\right) -
    (1-x)\left(1 - \frac{Q^2_\mathrm{min}}{Q^2_\mathrm{max}}\right) \right] \,.
    \label{eq:proton-flux-approx}
\end{equation}

\paragraph{Dipole form factor}

For the proton, light ions or as a simple model for extended particles, we use a generic dipole form factor:

\begin{equation}
    F_\mathrm{dipole} (Q^2) = \frac{1}{\left(1+Q^2/Q_0^2\right)^2}
\end{equation}

The scale $Q_0^2$ is a free parameter and is set to $0.71\ \GeV^2$.

Another baseline is given by the simple approximation of small virtualities, $Q^2 \to 0$, hence

\begin{equation}
    F_\mathrm{dipole,approx} (Q^2) = 1
\end{equation}

\paragraph{Gaussian form factor}

As an alternative phenomenological parametrization, we implement a Gaussian form factor:

\begin{equation}
    F_\mathrm{gauss} (Q^2) = \exp \left( - \frac{Q^2}{2 Q_0^2}\right)
\end{equation}

The Gaussian form factor provides a steeper falloff at large $Q^2$ compared to the dipole.

\paragraph{Homogeneously-charged sphere form factor}

This distribution assumes a uniform charge density within a sphere of radius $R$ and zero outside.
The corresponding form factor is

\begin{equation}
    F_\mathrm{HCS} (Q^2) = \frac{3}{(QR)^3} \left[\sin(QR) - QR \cos(QR)\right]
    \label{eq:hsc-ff}
\end{equation}

where $Q = \sqrt{Q^2}$ and $R$ is the nuclear radius.
This model is overly simplistic for realistic nuclei, and while it can in principle be used,
we omit it for the phenomenological studies in this paper.

\paragraph{Woods-Saxon form factor}

For heavy ions such as lead or gold, we use the Woods-Saxon nuclear density distribution:

\begin{equation}
    \rho(r) = \frac{\rho_0}{1 + \exp\left((r - R_\mathrm{WS})/d\right)}
    \label{eq:woods-saxon-density}
\end{equation}

Here, $R_\mathrm{WS}$ is the effective radius, and $d$ is the skin depth parameter.
Typical values are $d \approx 0.55$ fm for most heavy nuclei,
and the radius $R_\mathrm{WS}$ scales as $R_\mathrm{WS} = r_0 \, A^{1/3}$ fm~\cite{Dudek:1981zz}.
We fix $r_0 = 1.18$ fm from the measured value $R_\mathrm{WS} = 6.624$ fm for $^{208}$Pb~\cite{DeJager:1974liz,DeVries:1987atn}.
The form factor is computed as the Fourier transform of the density:

\begin{equation}
    F_\mathrm{WS} (Q^2) = 4\pi \int_0^\infty \frac{\sin(Qr)}{Q} \, \rho(r) r \, \mathrm{d}r
    \label{eq:woods-saxon-ff}
\end{equation}

The normalization constant $\rho_0$ is fixed such that $F(Q^2 = 0\ \GeV^2) = 1$.
This integral must be evaluated numerically,
and we pre-compute a lookup table of $F_\mathrm{WS}(Q^2)$ on a logarithmic grid spanning $Q^2 \in [10^{-12}, 10^4]$ GeV$^2$ with $10^5$ points.

\paragraph{Approximate Woods-Saxon form factor}

We also implement the approximate form from Ref.~\cite{Klein:1999qj},
which treats the Woods-Saxon distribution as a hard sphere convolved with a Yukawa potential:

\begin{equation}
    F_\mathrm{WS, approx} (Q^2) = \frac{3}{(QR_\mathrm{WS})^3} \left[\sin(QR_\mathrm{WS}) - QR_\mathrm{WS} \cos(QR_\mathrm{WS})\right] \frac{1}{1 + a^2 Q^2}
    \label{eq:woods-saxon-approx}
\end{equation}

where $a \approx 0.7$ fm characterizes the range of the Yukawa tail.
This approximation captures the features of the Woods-Saxon distribution while remaining analytic.

\begin{figure}[ht]
\centering
\begin{tabular}{cc}
    \includegraphics{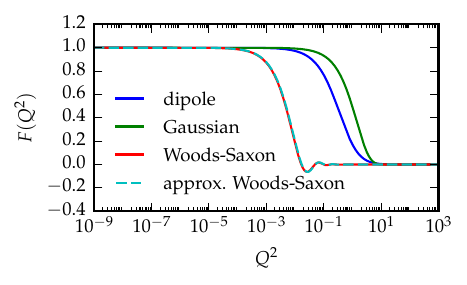} &
    \includegraphics{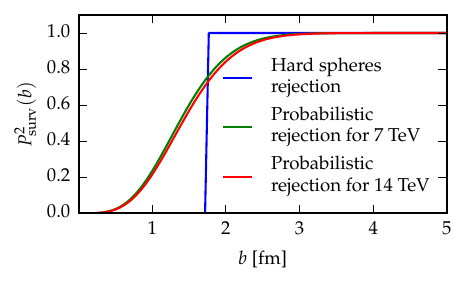}
\end{tabular}
\caption{Left: Different form factor parametrisations $F(Q^2)$.
The dipole and Gaussian form factors have been evaluated at parameters appropriate for protons,
the (approximated) Woods-Saxon for lead.
The latter two lie on top of each other and are hence indistinguishable.
Right: Comparison of hard spheres and probabilistic rejection, at $\sqrt{s} = 7$ and 14 TeV, for proton beams using $R_p = 4.45\ \GeV^{-1} = 0.88$~fm.
}\label{fig:ff-comparison}
\end{figure}

In the left panel in Fig.~\ref{fig:ff-comparison} we compare the different form factors,
inserting where necessary appropriate parameters for proton and lead beams.
The approximate formula for the Woods-Saxon yields in fact very similar behaviour
as the true Woods-Saxon potential.

\begin{figure}[ht]
\centering
\begin{tabular}{cc}
    \includegraphics{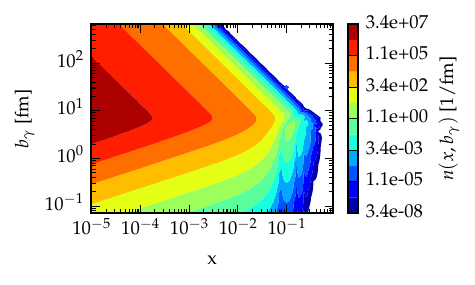} &
    \includegraphics{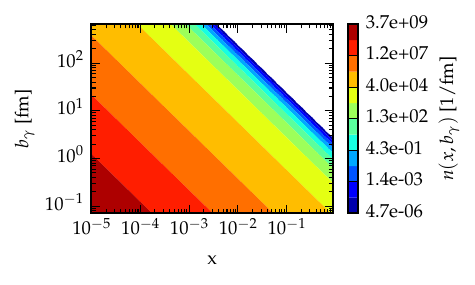}
\end{tabular}
\caption{Photon flux $n(x, b_\gamma)$ for the Woods-Saxon form factor, Eq.~\ref{eq:woods-saxon-ff},
and in the point-like approximation, Eq.~\ref{eq:point-like-flux}.
The range has been restricted to 15 orders of magnitude for clarity. }\label{fig:n_x_b}
\end{figure}

For computations with extended form factors,
we transition to the point-like approximation for large impact parameters, similar to Ref.~\cite{Eskola:2024fhf}.
In this region, the photon flux behaves like $\exp(-x b)$ in the limit $b \to \infty$~\cite{Krauss:1997vr}.
Typically, $b_\mathrm{threshold} \sim (5\text{--}10) \, R$, where $R$ is the particle radius.
This transition exploits the exponential suppression and avoids excessive need for the numerical integration,
hence accelerating the calculation, while maintaining exact treatment of the
relevant region of close encounters in the collision kinematics.

\subsection{Impact-parameter sampling}

We sample the impact parameter $b_\gamma$
from a distribution $p(b_\gamma)$ chosen to efficiently cover the kinematically relevant region.
We use

\begin{equation}
    p(\tilde{b}) = \frac{1}{\mathcal{N}} \frac{\tilde{b}}{\tilde{b}^2 + 1} \quad \text{for} \quad \tilde{b} \in [\tilde{b}_\mathrm{min}, \tilde{b}_\mathrm{max}]
    \label{eq:b-sampling}
\end{equation}

where $\tilde{b} \equiv b_\gamma/R$ is the impact parameter in units of the particle radius $R$,
and $\mathcal{N}$ is a normalization constant.
This distribution provides better efficiency than a uniform or log-normal distribution, matching the approximate power-law behavior of the flux at intermediate $b$ and the peak at $b = R$.

While for the minimum any value $\tilde{b}_\mathrm{min} \le 1$ can be justified,
$\tilde{b}_\mathrm{max}$ has to be large enough not to cut off phenomenologically relevant parts of the phase space for small $x$.

\subsection{Overlap rejection}

To arrive at physically meaningful predictions,
we must exclude configurations where the colliding particles overlap,
to suppress configurations where strong interactions occur additionally
and hence would experimentally not be classified as elastic exclusive production.
We consider two prescriptions for this survival factor $P_\mathrm{surv}^2 (b)$.

The impact parameter as the distance between the two incoming beams is calculated
from the individual photon impact parameters $b_\gamma$ by the law of cosine, as

\begin{equation}
    b = \sqrt{b_{1,\gamma}^2 + b_{2,\gamma}^2 - 2 b_{1,\gamma} b_{2,\gamma} \cos \phi}
\end{equation}

with $\phi$ the angle between the two impact parameter vectors in the 2D plane.

\paragraph{Hard-sphere rejection}

The simplest approach treats the particles as hard spheres of radii $R_1$ and $R_2$.
Events are rejected if $b < R_1 + R_2$, leading to a survival factor

\begin{equation}
    P_\mathrm{surv}^2 (b) = \Theta(b - R_1 - R_2)
\end{equation}

It has been shown to already provide a very good description for nuclear collisions.
Since inelastic interactions would dominate the photon flux in this region, we use only this hard-sphere rejection.

\paragraph{Probabilistic survival (proton-proton)}

For proton-proton collisions, this picture can be refined for close-encounter collisions
by parametrising the probability with an informed fit to data.
Following Ref.~\cite{Dyndal:2014yea}, we implement a probabilistic survival factor:

\begin{equation}
    P_\mathrm{surv}^2 (b) = \left| 1 - \exp\left(-\frac{b^2}{2 b_0(s)}\right) \right|^2
    \label{eq:survival-prob}
\end{equation}

This parametrises the probability that the two protons do not undergo an inelastic collision.
The scale $b_0(s)$ depends on the center-of-mass energy $\sqrt{s}$ and is fitted to data.
We use the parametrization from Ref.~\cite{Shao:2022cly}:

\begin{equation}
    b_0(s) = A + B \, \log(s/\GeV^2) + C \, \log^2(s/\GeV^2)
\end{equation}

with $A = 9.81$ GeV$^{-2}$, $B = 0.211$ GeV$^{-2}$, and $C = 0.0185$ GeV$^{-2}$.
This fit describes \LHC data and interpolates smoothly to lower energies.

As required, the survival probability approaches unity for $b \gg \sqrt{b_0}$, and falls to zero as $b \to 0$.
For typical \LHC energies, $\sqrt{s} = 13$ TeV, $b_0 \approx 20$ GeV$^{-2}$,
corresponding to a characteristic size $\sqrt{b_0} = 4.5$ GeV$^{-1} = 0.9$ fm.

We compare these two rejection models in the right panel of Fig.~\ref{fig:ff-comparison} for proton beams.

\subsection{Factorization formula}

The total cross-section for the process $AA \to AXA$,
where both nuclei $A$ remain intact and produce a system $X$ via photon fusion, is given by

\begin{equation}
    \sigma_{AA \to AXA} = \int \mathrm{d}x_1 \int \mathrm{d}x_2
    \int \mathrm{d}^2\mathbf{b}_1 \int \mathrm{d}^2\mathbf{b}_2
    \, n(x_1, |\mathbf{b}_1|) \, n(x_2, |\mathbf{b}_2|) \, P_\mathrm{surv}^2(|\mathbf{b}_1-\mathbf{b}_2|)
    \int \mathrm{d}\Phi_X \, \frac{\mathrm{d}\hat{\sigma}_{\gamma \gamma \to X}}{\mathrm{d}\Phi_X}
    \label{eq:factorization-full}
\end{equation}

where $n(x, b)$ is the photon flux from Eq.~\ref{eq:fourier-transform},
$\mathbf{b}_{1,2}$ are the impact-parameter vectors of the two colliding particles,
$P_\mathrm{surv}^2$ is the survival probability from Eq.~\ref{eq:survival-prob},
and where $\Phi_X$ denotes the phase space of the final state $X$ and the differential cross-section.

For photon fluxes in which the impact parameter has been integrated out, this simplifies to

\begin{equation}
    \sigma_{AA \to AXA} = \int \mathrm{d}x_1 \int \mathrm{d}x_2
    \, n(x_1) \, n(x_2) \,
    \int \mathrm{d}\Phi_X \, \frac{\mathrm{d}\hat{\sigma}_{\gamma \gamma \to X}}{\mathrm{d}\Phi_X} \ .
    \label{eq:factorization-simple}
\end{equation}

\section{Comparison to data for \texorpdfstring{$pp$}{pp} and Pb+Pb beams}\label{Sec:Validation}

Having provided all necessary equations that underpin our calculation, in this section we compare its results with experimental data.
The calculation relies on a pre-release version of \Sherpa v3.1~\cite{Sherpa:2024mfk},
using the tree-level matrix element generators \Comix and \Amegic~\cite{Gleisberg:2008fv,Krauss:2001iv}.
We explore the impact of higher-order corrections through dedicated runs with soft-photon resummation using
Yennie-Frautschi-Suura formalism~\cite{Yennie:1961ad,Schonherr:2008av} and with
Next-to-Leading Order (NLO) in the electroweak coupling using
local subtraction with Catani-Seymour dipoles~\cite{Catani:1996vz,Gleisberg:2007md,Schonherr:2017qcj}.

Two measurements of exclusive dimuon production,
$\gamma\gamma\to\mu^+\mu^-$, are used for validation.
The first, by \ATLAS in $pp$ collisions at $\sqrt{s}=13$~TeV~\cite{ATLAS:2017sfe},
measured the dimuon invariant mass
in the range $12 < m_{\mu\mu} < 70$~GeV.
The second, by \ATLAS in ultra-peripheral Pb+Pb collisions at
$\sqrt{s_{\rm NN}}=5.02$~TeV~\cite{ATLAS:2020epq},
measured double-differential distributions of dimuon rapidity $|y_{\mu\mu}|$,
scattering angle $\cos\theta^*$, and dimuon invariant mass $m_{\mu\mu}$.
The fiducial cuts applied in both analyses are summarised
in Table~\ref{tab:cuts}.
We compare to available data using \Rivet~\cite{Bierlich:2024vqo,Buckley:2019xhk}.

\begin{table}[htbp]
  \centering
  \begin{tabular}{lcc}
    \toprule
    Observable / Cut
      & $pp$ at $\sqrt{s}=13$~TeV~\cite{ATLAS:2017sfe}
      & Pb+Pb at $\sqrt{s_{\rm NN}}=5.02$~TeV~\cite{ATLAS:2020epq} \\
    \midrule
    Muon pseudorapidity          & $|\eta_\mu|<2.4$                          & $|\eta_\mu|<2.4$ \\
    Muon $p_T$                   & $>6$~GeV, $>10$~GeV if $m_{\mu\mu}>30$~GeV & $>4$~GeV \\
    Dimuon invariant mass        & $12 < m_{\mu\mu} < 70$~GeV              & $m_{\mu\mu}>10$~GeV \\
    Dimuon $p_T$   & ---                                       & $p_{T,\mu\mu}<2$~GeV \\
    \bottomrule
  \end{tabular}
  \caption{Fiducial cuts in the two \ATLAS measurements used for validation.}\label{tab:cuts}
\end{table}

\subsection{Comparing form factors and assessing resulting uncertainties}\label{subsec:formfactors}

\begin{figure}[htpb]
  \centering
  \begin{tabular}{cc}
    \includegraphics[width=0.49\linewidth]{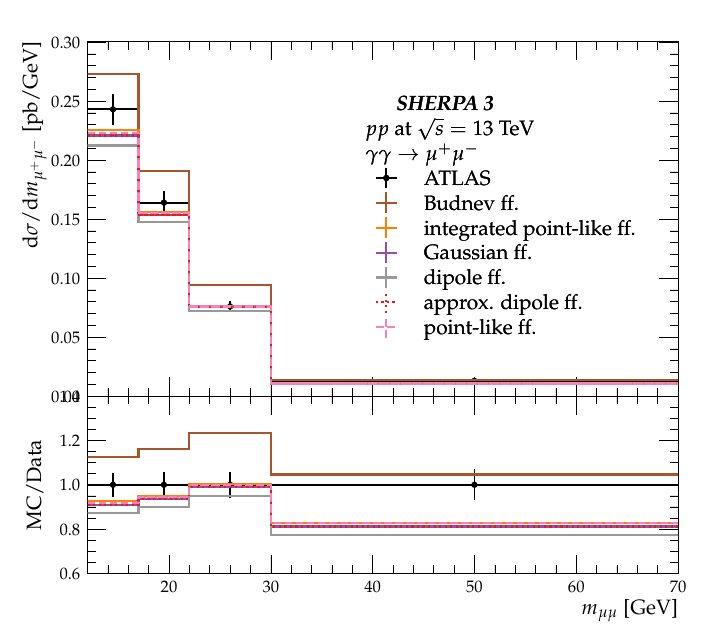} &
    \includegraphics[width=0.49\linewidth]{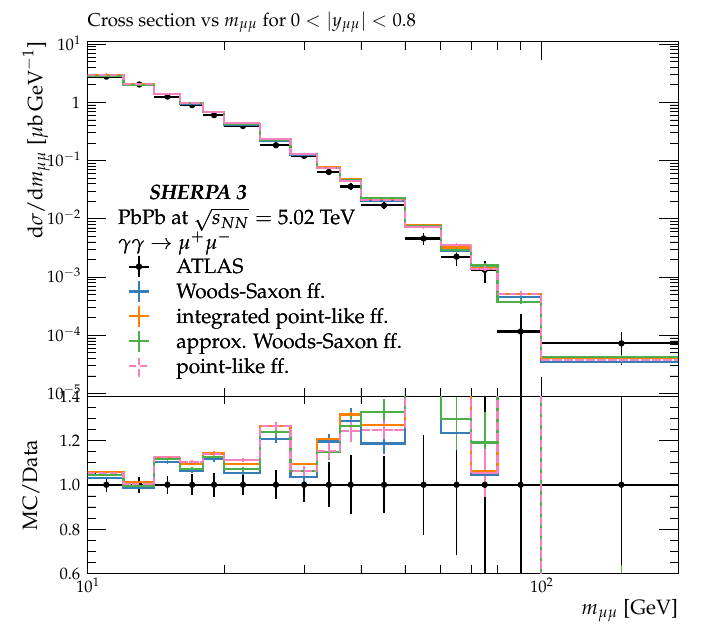}
  \end{tabular}
  \caption{Left: Differential dimuon invariant mass
  at LO comparing different proton form factors, data measured by \protect\ATLAS
  in $pp$ at $\sqrt{s}=13$~TeV~\protect\cite{ATLAS:2017sfe}.
  Right: Differential dimuon invariant mass for rapidities $0 < | y_{\mu\mu} | < 0.8$ at LO comparing different ion form factors,
  data measured by \protect\ATLAS in UPC Pb+Pb at
  $\sqrt{s_{\rm NN}}=5.02$~TeV~\protect\cite{ATLAS:2020epq}.}\label{fig:pp:compare-ffs}
\end{figure}

In Fig.~\ref{fig:pp:compare-ffs}, we compare predictions for the
invariant mass spectrum $m_{\mu^+\mu^-}$ of exclusively produced
dimuon pairs to data from \ATLAS~\cite{ATLAS:2017sfe} in $pp$ collisions at
$\sqrt{s} = 13$~TeV (left) and to UPC Pb+Pb data from
\ATLAS~\cite{ATLAS:2020epq} at $\sqrt{s_{\rm NN}} = 5.02$~TeV (right).
For proton beams we consider the Budnev, dipole, Gaussian, approximate dipole,
and the two variants of the point-like form factor.
To isolate effects of the different form factors, we do not yet apply any
overlap rejection and require a minimal emission distance $b_\gamma > R$ here.
With the Budnev parametrisation from~\cite{Budnev:1975poe}, the total
cross-section is too large by approximately 15\%.
The other form factors all undershoot the data, the dipole form factor by about 11\%
and the remaining ones by about 7\%. However, at this stage this shortfall in
cross-section is to be expected due to the limitation of the impact parameter,
and it highlights the importance of contributions at $b_\gamma < R$,
that we will elicit further in Sec.~\ref{subsec:bmin}.

For ion beams, we compare the Woods-Saxon (WS) form factor~\cite{DeVries:1987atn},
its analytic approximation, and the point-like limits, again restricted to runs
without any overlap rejection and with a naive cut $b_\gamma > R$.
The form factors are all in agreement with each other and describe the data well
for small dimuon masses. With increasing masses, however, the calculations yield
higher cross sections than measured. This can easily be explained by the fact
that this region probes large values of $x$. As seen in
Fig.~\ref{fig:n_x_b}, the flux in this region drops off sharply with increasing
$b_\gamma$, and hence effects at $b_\gamma \approx R$ become important.

For the remainder of this work, we will only consider the dipole form factor for
proton beams and the Woods-Saxon form factor for Pb beams.

\subsection{Impact of rejection modelling}\label{subsec:rejection}

\begin{figure}[htpb]
  \centering
  \begin{tabular}{cc}
    \includegraphics[width=0.49\linewidth]{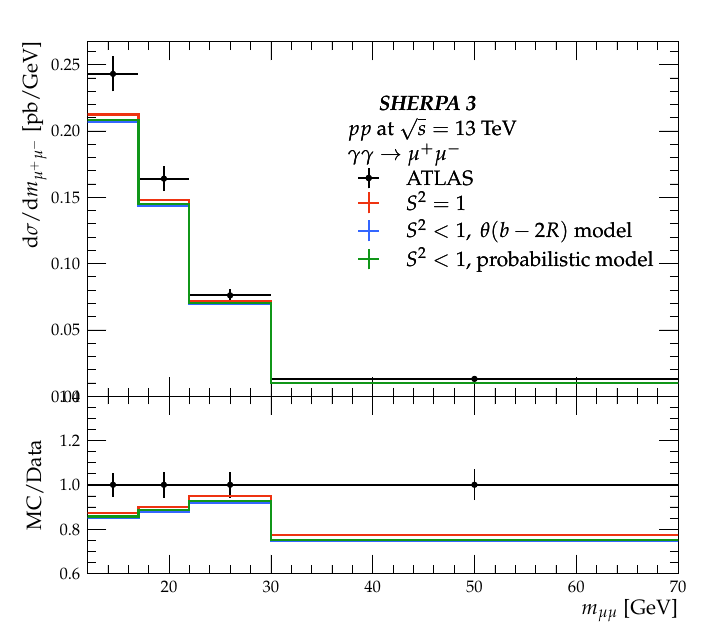} &
    \includegraphics[width=0.49\linewidth]{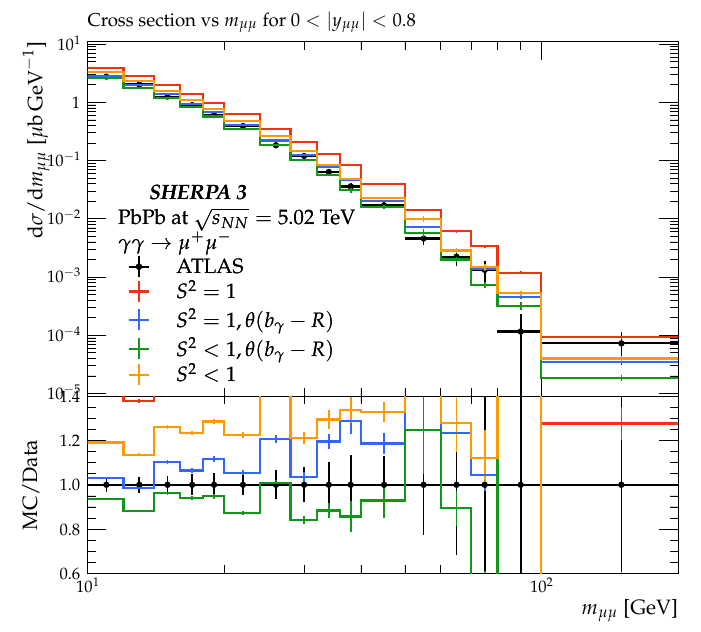}
  \end{tabular}
  \caption{Left: Differential dimuon invariant mass
  at LO with the dipole form factor, comparing different beam-overlap rejection models
  ($S^2=1$, $\theta(b{-}2R)$, probabilistic rejection),
  data measured by \protect\ATLAS in $pp$ at $\sqrt{s}=13$~TeV~\protect\cite{ATLAS:2017sfe}.
  Right: Differential dimuon invariant mass for rapidities $0 < | y_{\mu\mu} | < 0.8$
  at LO using the Woods-Saxon form factor, comparing the impact of overlap rejection
  and cuts on the emission distance,
  data measured by \protect\ATLAS in UPC Pb+Pb at
  $\sqrt{s_{\rm NN}}=5.02$~TeV~\protect\cite{ATLAS:2020epq}.}\label{fig:compare-rejection}
\end{figure}

The beam-overlap rejection suppresses contributions from impact parameters $b<2R$.
In this region, the two hadrons would overlap geometrically, and it is straightforward to assume that such configurations lead to inelastic collisions spoiling the exclusivity of the produced system.
We compare three approaches: no rejection, $S^2 = 1$; the hard-sphere, $\theta(b - 2R)$;
and additionally for protons the probabilistic model from Eq.~\ref{eq:survival-prob}.
In the left plot in Fig.~\ref{fig:compare-rejection}, we show the impact of these models
for the dipole form factor in $pp$ collisions, still restricted to $b_\gamma > R$.
The two rejection models decrease the cross section by about 3\% in this case.
Overall, the agreement is very good between the two in this measurement,
with the hard-sphere model rejecting just slightly more events than the
probabilistic model.

For ion beams, we only employ the hard-sphere rejection and on the right in
Fig.~\ref{fig:compare-rejection}, we compare four different calculations:
(i) no rejection and restricted impact parameters, $S^2 = 1,\ \theta(b_\gamma - R)$,
(ii) hard-sphere rejection and restricted impact parameters, $S^2 < 1,\ \theta(b_\gamma - R)$,
(iii) no rejection and unrestricted impact parameters, $S^2 = 1$,
and (iv) hard-sphere rejection and unrestricted impact parameters, $S^2 < 1$.

These cuts have a strong impact on the cross section. On the one hand,
when comparing the effect of the $b_\gamma$ cut, we see that the phase
space $b_\gamma < R$ adds an additional 50\% and 30\%, for the cases of
$S^2 = 1$ and $S^2 < 1$ respectively, to the cross section. On the other hand,
the vetoing of overlapping beams removes 10\% (when imposing $\theta(b_\gamma - R)$) and
30\% (without a $b_\gamma$ cut) of the cross section.
While including the full $b_\gamma$ phase space is clearly necessary,
the impact of the rejection modelling is naturally much more pronounced than
in the proton case and still subject to parameter choices, like the assumed nuclear radius.

The most realistic computation, including overlap rejection and including
the $b_\gamma < R$ phase space, predicts a cross section about 20\% larger than measured,
which we can partially attribute to missing higher-order effects.
We will discuss this in detail in Sec.~\ref{subsec:higher-order}.

\subsection{Impact of emission distance}\label{subsec:bmin}

\begin{figure}[htpb]
  \centering
  \begin{tabular}{cc}
    \includegraphics[width=0.49\linewidth]{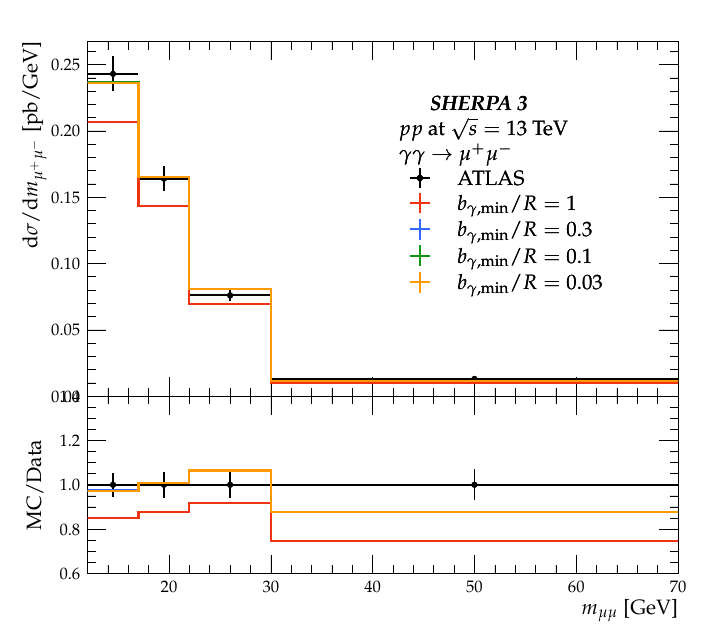} &
    \includegraphics[width=0.49\linewidth]{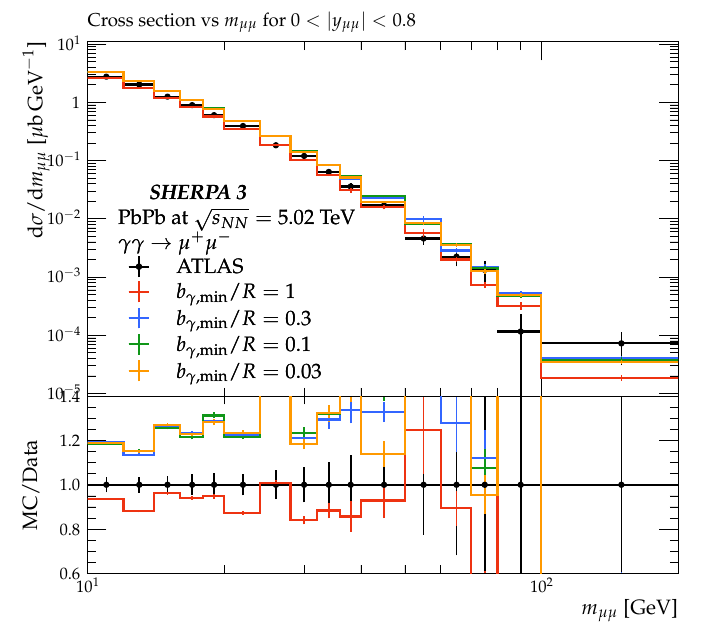}
  \end{tabular}
  \caption{Left: Differential dimuon invariant mass
  at LO with the dipole form factor, comparing predictions for
  $b_{\gamma,\mathrm{min}}/R \in \{0.03, 0.1, 0.3, 1\}$,
  data measured by \protect\ATLAS in $pp$ at $\sqrt{s}=13$~TeV~\protect\cite{ATLAS:2017sfe}.
  Right: Same comparison with the Woods-Saxon form factor,
  using differential dimuon invariant mass for
  rapidities $0 < | y_{\mu\mu} | < 0.8$ measured by \protect\ATLAS in UPC Pb+Pb at
  $\sqrt{s_{\rm NN}}=5.02$~TeV~\protect\cite{ATLAS:2020epq}.}\label{fig:compare-bmin}
\end{figure}

While large overlap between the emitters is implausible and must be rejected,
the emission of the photon can in principle happen at small distances from the
emitter, even within the emitter radius in the transverse plane.
Often, a minimal impact parameter $b_{\gamma,\mathrm{min}} = R$ is imposed,
just like in our comparisons so far.
We now relax this constraint and present in Fig.~\ref{fig:compare-bmin}
computations for a range of $b_{\gamma,\mathrm{min}} < 1$ values for both proton (left)
and lead beams (right).

Clearly, the restriction of $b_{\gamma,\mathrm{min}} = R$ is too simplistic and lacks
significant contributions to the phase space. Lifting this constraint we see
an increase of the cross section of about 15\%. We also observe
that the calculation converges quickly: a value of
$b_{\gamma,\mathrm{min}}/R = 0.3$ already seems to capture all relevant contributions,
and lowering the cutoff further has a negligible impact on the predictions.
The same behaviour is observed for the ion case, where the convergence
sets in at the same fraction of the nuclear radius.
This demonstrates that contributions from within the nuclear radius are
physically present and must not be excluded by an artificial cutoff.
The convergence of the predictions justifies a default value of $b_{\gamma,\mathrm{min}}/R = 0.3$.

\subsection{Impact of Next-to-Leading Order in QED and soft-photon resummation}\label{subsec:higher-order}

\begin{figure}[htpb]
  \centering
  \begin{tabular}{cc}
    \includegraphics[width=0.49\linewidth]{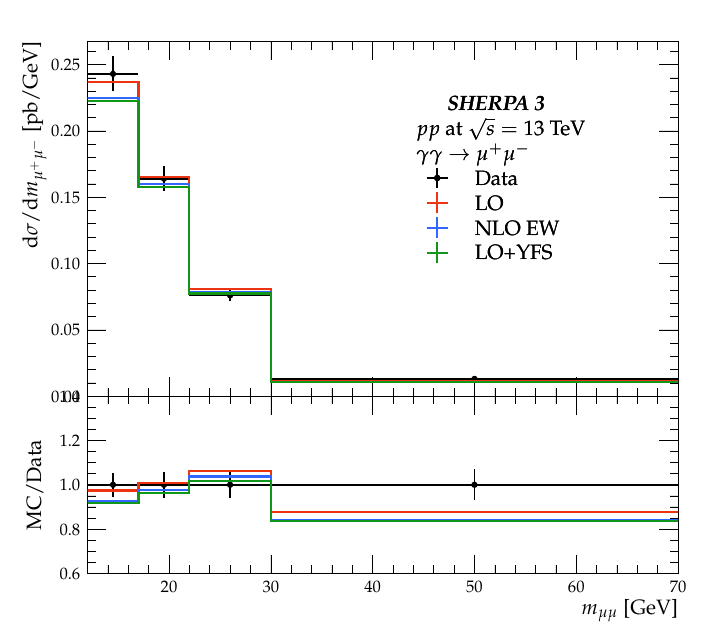} &
    \includegraphics[width=0.49\linewidth]{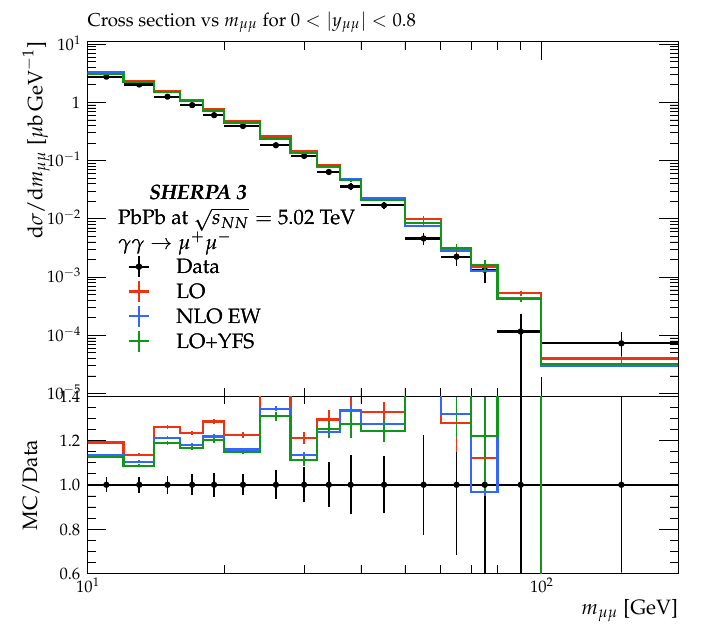}
  \end{tabular}
  \caption{Left: Differential dimuon invariant mass
  at LO, NLO EW, and LO+YFS with the dipole form factor,
  data measured by \protect\ATLAS in $pp$ at $\sqrt{s}=13$~TeV~\protect\cite{ATLAS:2017sfe}.
  Right: Differential dimuon invariant mass for rapidities $0 < | y_{\mu\mu} | < 0.8$ at LO, NLO EW, and LO+YFS with the Woods-Saxon form factor,
  data measured by \protect\ATLAS in UPC Pb+Pb at
  $\sqrt{s_{\rm NN}}=5.02$~TeV~\protect\cite{ATLAS:2020epq}.}\label{fig:compare-accuracies}
\end{figure}

In Sec.~\ref{subsec:rejection} we observed that the most realistic
leading-order calculation overpredicts the cross section by
approximately 20\%, which we attributed partially to missing higher-order effects.
To quantify this contribution, we compare in Fig.~\ref{fig:compare-accuracies}
three calculations: leading-order (LO); Next-to-Leading Order electroweak
corrections (NLO EW); and LO with soft-photon resummation (LO+YFS).
All calculations use the dipole form factor for $pp$ collisions and
the Woods-Saxon form factor for Pb+Pb, with $b_{\gamma,\mathrm{min}}/R = 0.3$
and hard-sphere rejection.

Both NLO EW and LO+YFS reduce the predicted cross section relative to LO.
On the left in Fig.~\ref{fig:compare-accuracies}, for the $pp$ measurement at
$\sqrt{s}=13$~TeV, the reduction in the lowest mass bin reaches about 10\%.
At higher masses, the corrections remain at a few percent.
The NLO EW prediction is about 3\% lower than LO+YFS at low mass, and the
two converge at higher masses.
The same pattern is observed in the Pb+Pb collisions at
$\sqrt{s_{\rm NN}}=5.02$~TeV (on the right), with similar magnitudes.

The large size of the corrections arises because the analyses were undertaken
using bare leptons, which are not infrared-safe observables.
In our calculation we yield infrared-finite results by using massive leptons.
Still, soft and collinear photon emission from the final-state leptons reduce the observed
transverse momentum and invariant mass of the leptons, leading to bin migration
from higher to lower bins and sensitivity to phase-space cuts.

One can observe in both analyses that the resummed photon emissions in
the YFS formalism leads to a slightly stronger reduction
in the differential cross-section as compared to the effects
from the NLO EW which effectively only captures the first emission.

Although the inclusion of higher-order corrections partially resolves
the discrepancy in the Pb+Pb data, c.f.\ Sec.~\ref{subsec:rejection},
reducing the overprediction from about 20\% to about 10\%, an overall offset remains.
This behaviour has been observed before and was critically examined in Ref.~\cite{Harland-Lang:2021ysd}.
In that work, different modelling uncertainties were examined like the form factor, survival probability,
radiative effects and implicit cuts on the impact parameter $b_\gamma$ that are applied in some calculations;
however, the source of the discrepancy of 10\% remained unclear.

Our computation takes into account radiative effects by means of the NLO corrections and the YFS resummation respectively,
and does include full dependence of the impact parameter including the survival probability.
As seen in Sec.~\ref{subsec:formfactors} different ion form factors agree with each other
within a few percent.
All this is confirming the findings from Ref.~\cite{Harland-Lang:2021ysd} about how these effects decrease
but do not resolve the difference between theory and experiment.

Very recently, this was investigated further in Ref.~\cite{Dyndal:2026uvm}.
Their study showed that a significant fraction of around 10\% of events was rejected
by the veto on central activity in the detector, through hadron production from accompanying
electromagnetic dissociation of the nuclei.
The analysis was inclusive on the ion dissociation, i.e.\ not requiring the nucleus to stay intact.
However, as was shown in Ref.~\cite{Dyndal:2026uvm}, the number of hadrons being emitted into
the central rapidity region violating the no-activity veto might have been underestimated,
hence leading to a measured cross section being lower than predicted.

\section{Predictions for photon-induced processes}
\label{Sec:Predictions}

The framework allows for arbitrary final states,
including extensions of the Standard Model through the UFO interface.
As two examples we present phenomenological studies of two processes,
both of which were measured or are in the process of being analysed.

\subsection{Off-shell exclusive \texorpdfstring{$W^+W^-$}{WW} production in \texorpdfstring{$pp$}{pp} collisions}\label{subsec:ww}

In Ref.~\cite{ATLAS:2016lse}, the exclusive production of $W$-boson pairs was
measured in the leptonic channels, i.e.\ $\gamma \gamma \to W^+ W^- \to e^\pm \mu^\mp \nu \bar{\nu}$.
This production mechanism is especially interesting as a direct probe of quartic gauge couplings
in the Standard Model, where the two-photon interaction probes the $\gamma\gamma WW$ vertex.
As discussed for the tau-pair production in~\cite{Dittmaier:2025ikh} one needs to
use a mixed EW scheme to avoid large logarithmic corrections arising
from the running of $\alpha$ from $Q=0$ to the electroweak scale,
which would appear in the $G_\mu$ scheme.
We compute the LO, LO+YFS and NLO EW corrections in the mixed scheme by rescaling
the cross section involving $l$ external photons with a factor $(\alpha(0)/\alpha_{G_\mu})^l$.

The analysis reported the total cross section but no differential data.
The measurement was requiring $p_{T,l} > 25$ and 20~GeV for the leading and subleading lepton, respectively,
and the lepton pairs had to additionally fulfill $m_{e\mu} > 20$~GeV for
the dilepton mass and $p_\mathrm{T}^{e\mu} > 30$~GeV for the dilepton transverse momentum.
However, the final cross section was extrapolated to full phase space using the
acceptance efficiency and the exclusivity efficiency and including the dissociation component, yielding a combined factor of $31.9$.
We undo this extrapolation of the total cross section and compare only within the fiducial region.
As the propagation of errors is not clear, we only rescale it, but do not unfold the different components.

Also, similar to the analyses from Sec.~\ref{Sec:Validation},
this analysis used bare leptons.
In our calculation we however choose to apply a dressing algorithm with $R =1$
to avoid divergent behaviour.

In Tab.~\ref{tab:WW-xs} we present the cross sections for this process.
The calculations are compatible with the measured value well within the large experimental uncertainties.
As discussed in Sec.~\ref{subsec:higher-order}, we see the YFS and NLO corrections
reducing the observed cross section by 3.4\% and 1.3\%, respectively, as a consequence of the bare leptons in the analysis.

\begin{table}[htbp]
  \centering
  \begin{tabular}{lcccc}
    \toprule
    & \ATLAS~\protect\cite{ATLAS:2016lse} & LO & LO+YFS & NLO EW \\
    \midrule
    $\sigma\left(\gamma \gamma \to W^+ W^-\right)\ [\mathrm{fb}]$ &
    $0.22(8)$ &
    $0.1846(11)$ &
    $0.17829(9)$ & 
    $0.18216(6)$ \\ 
    \bottomrule
  \end{tabular}
  \caption{Cross sections, in fb, for $\gamma \gamma \to e^\pm \mu^\mp \nu \bar{\nu}$ for $pp$ beams at $\sqrt{s} = 8$ TeV at the \LHC.}\label{tab:WW-xs}
\end{table}

\begin{figure}[htpb]
  \centering
  \begin{tabular}{cc}
    \includegraphics[width=0.49\linewidth]{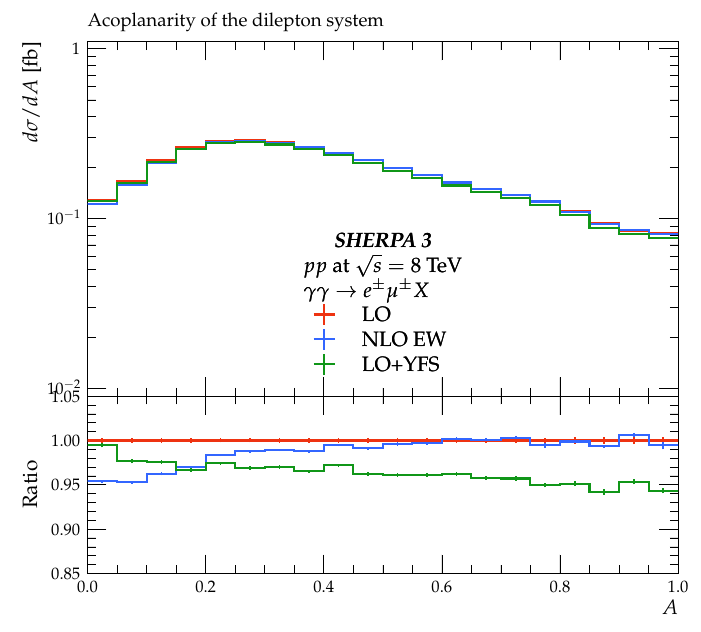} &
    \includegraphics[width=0.49\linewidth]{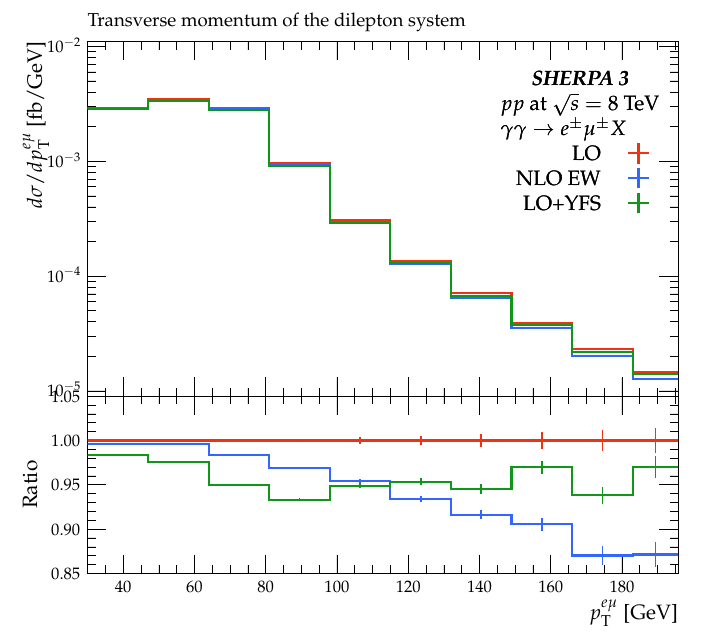}
  \end{tabular}
  \caption{Differential distribution of acoplanarity (left) and dilepton
  transverse momentum, $p_\mathrm{T}^{e\mu}$, (right) for exclusive $W^+W^-$ production as
  measured in~\protect\ATLAS~\protect\cite{ATLAS:2016lse}, comparing LO, LO+YFS and NLO predictions. }\label{fig:ww-xs}
\end{figure}

We present differential predictions of the LO,
LO+YFS and NLO EW computations in Fig.~\ref{fig:ww-xs}.
On the left panel, we show the distribution of acoplanarity $A = 1 - |\Delta\phi|/\pi$ of the dileptons and
on the right panel the dilepton transverse momentum $p_\mathrm{T}^{e\mu}$.
One can observe that the NLO corrections and the soft-photon resummation differ in size and shape for this measurement.
The NLO correction is largest for large transverse momenta and in the back-to-back region,
reaching about -15\% and -5\%, respectively.
Conversely it is converging to the LO prediction for small transverse momenta and small dilepton angles.
The YFS resummation yields a rather flat correction; interestingly, the correction tends towards the LO for the back-to-back region and small transverse momenta.

\subsection{Exclusive \texorpdfstring{$\tau$}{tau}-pair production in Pb+Pb and $pp$ collisions}

In a recent \ATLAS measurement~\cite{ATLAS:2026wrz}, exclusive $\tau$-lepton pair production
via the two-photon process $\gamma\gamma \to \tau^+\tau^-$ was studied in ultra-peripheral
Pb+Pb collisions at $\sqrt{s_{NN}} = 5.02$~TeV.
Beyond providing a clean test of QED at high invariant mass, such measurements have become
a competitive probe of the $\tau$-lepton anomalous
magnetic moment $a_\tau = (g-2)_\tau/2$~\cite{ATLAS:2022ryk,CMS:2022arf,CMS:2024qjo}.
A reliable extraction of $a_\tau$ necessitates precise Standard Model predictions for the
signal, including higher-order corrections.
We make predictions for that measurement as well as for a prospective analogous analysis for $pp$ collisions.

Three signal regions are defined based on the reconstructed final state:
(i) one muon plus one track not associated with the muon, 1M1T,
(ii) one muon plus three tracks not associated with the muon, 1M3T, and
(iii) one muon plus one electron, 1M1E.
All signal leptons are required to have $p_T > 4$~GeV and $|\eta| < 2.5$, while tracks
must satisfy $p_T > 0.1$~GeV and $|\eta| < 2.5$.
Association of tracks and leptons is performed using a cone size of $\Delta R < 0.1$.
The 1M1T region imposes additional cuts on the system transverse momentum $p_T^\mathrm{sys} > 1$~GeV,
acoplanarity $A = 1 - |\Delta\phi|/\pi < 0.4$, and requires the leptonic track to have $p_T < 4$~GeV
with opposite charge to the muon.
The 1M3T region requires the invariant mass of the three-track system to be below 1.7~GeV,
acoplanarity below 0.2, and opposite total charge.
The 1M1E region imposes only an opposite-charge requirement and vetoes additional tracks.
Observables measured include the leading muon $p_T$, the $p_T$ of the track or electron,
the system transverse momentum, invariant mass, rapidity, pseudorapidity difference, and acoplanarity.

As before, we compute LO, LO+YFS and NLO~EW predictions with massive leptons
throughout and in the $\alpha(0)$ scheme.
The NLO~EW corrections apply to the $\gamma\gamma\to\tau^+\tau^-$ matrix
element only, whereas the YFS resummation is applied both to the $\tau$ pair and to
their decay products.
For the leptonic decays we additionally take NLO matrix-element corrections
to the decay into account.
All decays include the spin correlations to the decay vertex, except for one dedicated
run in which we switch off both the spin correlations and
the QED corrections, i.e.\ the YFS resummation (see below). We do not correct for breakups of the nuclei, necessary for the 0$n$0$n$ requirement applied in the analysis.

\begin{figure}[htpb]
  \centering
  \begin{tabular}{cc}
    \includegraphics[width=0.49\linewidth]{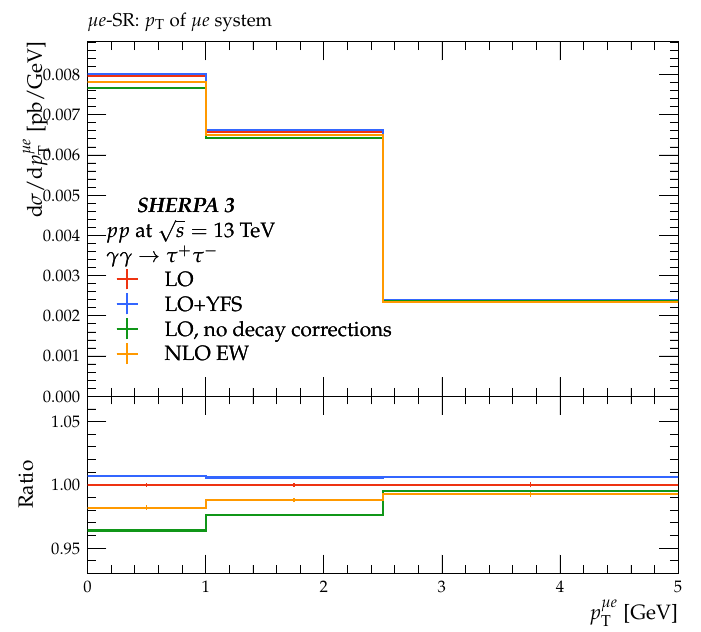} &
    \includegraphics[width=0.49\linewidth]{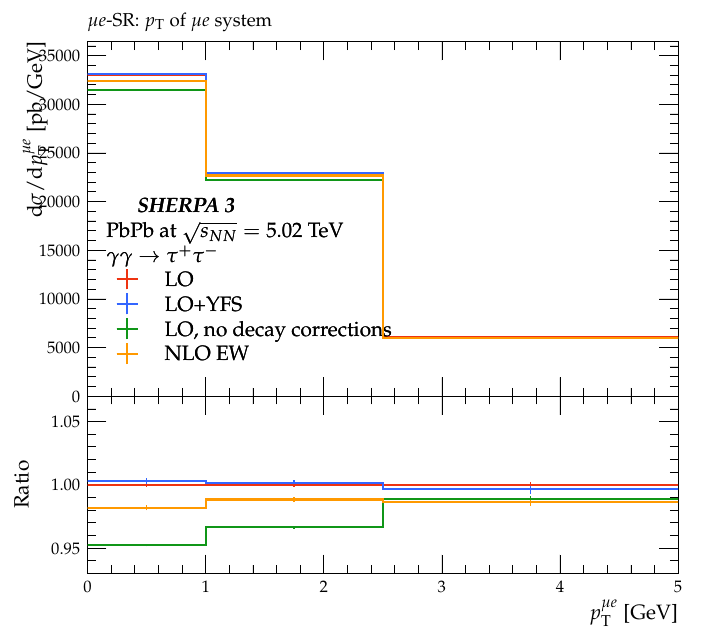} \\
    \includegraphics[width=0.49\linewidth]{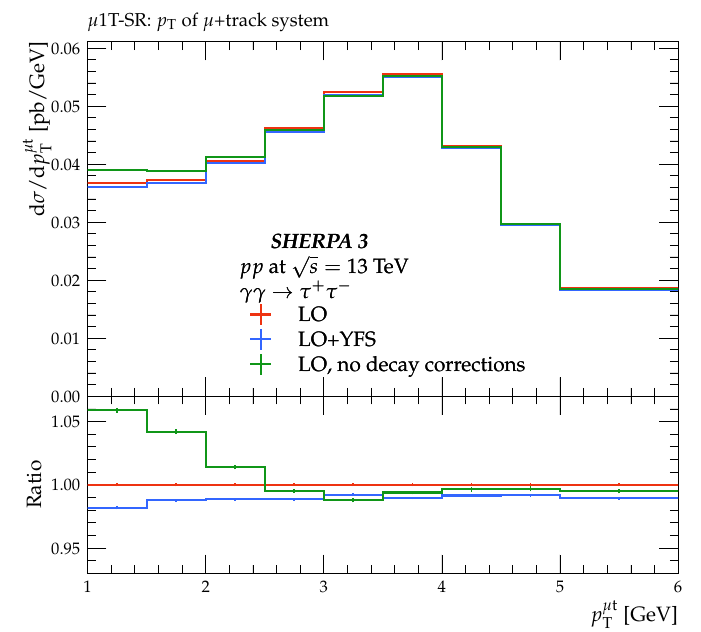} &
    \includegraphics[width=0.49\linewidth]{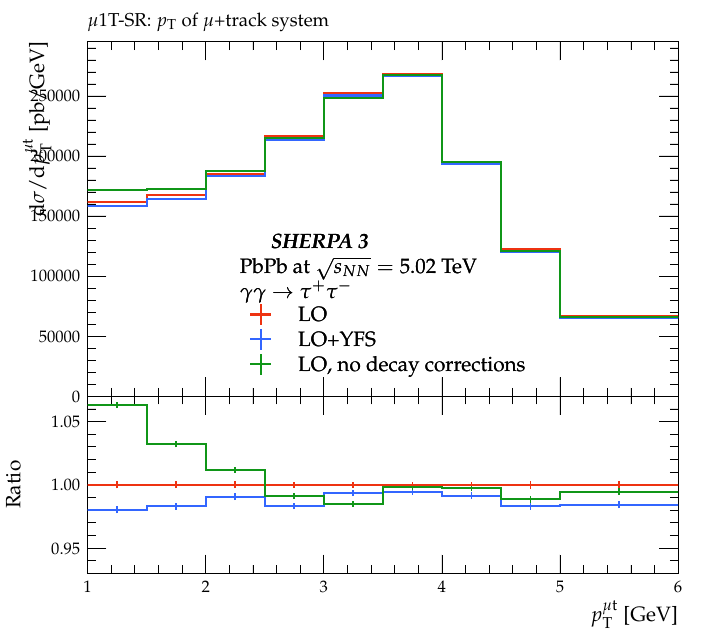} \\
    \includegraphics[width=0.49\linewidth]{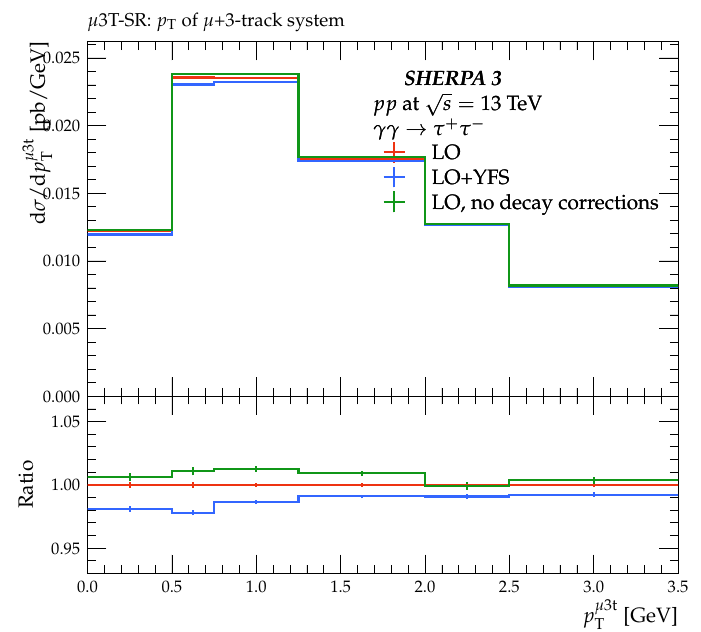} &
    \includegraphics[width=0.49\linewidth]{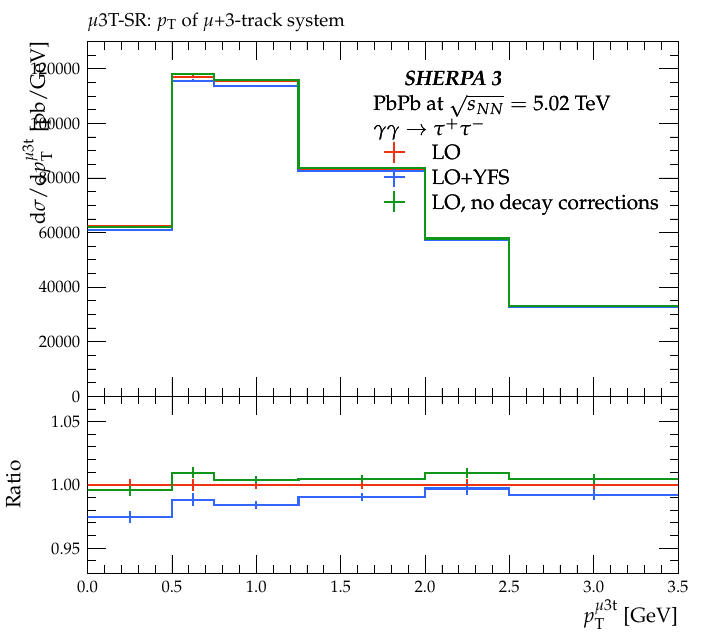}
  \end{tabular}
  \caption{Differential distribution of the system's transverse momentum,
  $p_\mathrm{T}^\mathrm{sys}$, for exclusive $\tau^+\tau^-$ production in $pp$ (left) and Pb+Pb
  (right) collisions at the \LHC, for the $\mu e$ (top), $\mu+$1track (middle) and $\mu+$3track (bottom) signal
  regions, comparing LO, LO+YFS and NLO~EW predictions. }\label{fig:tau-xs}
\end{figure}

In Fig.~\ref{fig:tau-xs} we present the visible-system transverse momentum $p_\mathrm{T}^\mathrm{sys}$
for the three signal regions, comparing $pp$ (left column) and Pb+Pb (right column) beams.
The two beam configurations yield almost identical shapes and relative corrections; the $pp$ cross
section is, however, roughly six orders of magnitude smaller, reflecting the coherent $Z^4$ enhancement
of the nuclear photon flux, only partially offset by the nuclear form factor at large photon energies.

Turning to the corrections, the spin correlations and the QED corrections at the decay
vertex produce large effects, but act in opposite directions for the leptonic and hadronic
channels: for the former they increase the cross section by up to 5\%, while for the latter
they decrease it by up to 7\%. In both cases the effect is most pronounced at small
$p_\mathrm{T}$. The soft-photon resummation alone changes the differential cross section at
the sub-percent level for the leptonic decays, but by up to 2\% for the hadronic ones.

For the $\mu e$ signal region, the NLO~EW corrections grow towards smaller $p_\mathrm{T}$ of
the decay system, reaching 2\%. This is effectively our best prediction, as it combines the
NLO~EW corrections to $\tau$-pair production with soft-photon resummation for the decays.

\section{Conclusion}
\label{Sec:Conclusion}

We have presented an automated framework for the simulation of photon-induced
processes in ultra-peripheral collisions of protons and nuclei, realised within
the \Sherpa event generator. Building on the equivalent-photon approximation, it
implements a comprehensive set of proton and nuclear form factor parametrisations
and retains the full dependence of the photon flux on the finite size of the
emitters and on their impact parameter. Configurations in which the colliding
particles overlap are removed through either a hard-sphere or a probabilistic
survival factor, and higher-order effects are included automatically, either by
resumming soft-photon emissions to all orders in the YFS formalism or through the
full NLO electroweak corrections. As the implementation is embedded in a
general-purpose event generator, it can leverage the automated interfaces,
for example to soft-photon resummation and decays with spin correlations,
and can be applied to arbitrary final states, including beyond the Standard Model
through the UFO interface.

We validated the framework against two \ATLAS measurements of exclusive dimuon
production, in $pp$ collisions at $\sqrt{s}=13$~TeV and in ultra-peripheral Pb+Pb
collisions at $\sqrt{s_{\rm NN}}=5.02$~TeV. The various form factors agree with one
another at the level of a few percent, so that their residual differences are
small compared to other sources of modelling uncertainty. We find that the photon
flux receives sizeable contributions from emission distances within the nuclear
radius, which must not be discarded by an artificial cutoff.
The inclusion of higher-order corrections lowers the predicted cross
section by up to 10\% in the lowest mass bins, with the YFS resummation and
the NLO electroweak calculation in good mutual agreement; their sizeable effect
reflects the use of bare leptons in the analyses, which are sensitive to soft and
collinear final-state radiation. While the $pp$ data are described well, a residual
offset of about 10\% remains in the Pb+Pb case, consistent with earlier
observations~\cite{Harland-Lang:2021ysd}.  This discrepancy has recently been attributed to electromagnetic
dissociation of the nuclei producing central hadronic activity that spoils the
exclusivity veto~\cite{Dyndal:2026uvm}.

As first applications, we presented predictions for off-shell exclusive $W$-pair
production in $pp$ collisions and for exclusive $\tau$-pair production in $pp$ and
Pb+Pb collisions. The predicted $W$-pair cross section is compatible with the
\ATLAS measurement within its still sizeable uncertainties, while the differential
electroweak corrections reach up to $-15\%$ in the kinematic tails of the distributions. For
$\tau$-pair production we provide predictions for the three experimental signal
regions in both $pp$ and Pb+Pb collisions, and find that the dominant corrections
originate from the spin correlations and QED radiation at the decay vertices, which
shift the cross section by up to $+5\%$ in the leptonic and $-7\%$ in the hadronic
channels, whereas the genuine NLO electroweak corrections to the production process
remain at the percent level. Our most complete prediction, combining these NLO
electroweak corrections with soft-photon resummation
to the decays, is provided for the $\mu e$ signal region.

The framework will be made available as part of a future \Sherpa v3.1 release.
Natural extensions of this work include the description of photonuclear and
photoproduction processes, such as jet production, and the application to
electron-ion collisions at the future \EIC, where the same formalism governs the
photon flux off the proton and the ion.

\section*{Acknowledgments}

We want to thank Marek Sch\"onherr for fruitful discussions and technical help during the course of this project.
FK acknowledges funding by STFC under grant agreement ST/P006744/1.
PM has been supported by the Swiss National Science Foundation (SNSF)
under contract 240015 and by the European Research Council (ERC)
under the European Union's Horizon 2020 research and innovation programme
grant agreement 101019620 (ERC Advanced Grant TOPUP).

\clearpage

\bibliographystyle{unsrt}
\bibliography{refs,spectra}

@article{CMS:2018erd,
    author = "Sirunyan, Albert M and others",
    collaboration = "CMS",
    title = "{Evidence for light-by-light scattering and searches for axion-like particles in ultraperipheral PbPb collisions at $\sqrt{s_\mathrm{NN}} =$ 5.02 TeV}",
    eprint = "1810.04602",
    archivePrefix = "arXiv",
    primaryClass = "hep-ex",
    reportNumber = "CMS-FSQ-16-012, CERN-EP-2018-271",
    doi = "10.1016/j.physletb.2019.134826",
    journal = "Phys. Lett. B",
    volume = "797",
    pages = "134826",
    year = "2019"
}

@article{ATLAS:2020hii,
    author = "Aad, Georges and others",
    collaboration = "ATLAS",
    title = "{Measurement of light-by-light scattering and search for axion-like particles with 2.2 nb$^{-1}$ of Pb+Pb data with the ATLAS detector}",
    eprint = "2008.05355",
    archivePrefix = "arXiv",
    primaryClass = "hep-ex",
    reportNumber = "CERN-EP-2020-135",
    doi = "10.1007/JHEP03(2021)243",
    journal = "JHEP",
    volume = "03",
    pages = "243",
    year = "2021",
    note = "[Erratum: JHEP 11, 050 (2021)]"
}

@article{CMS:2022arf,
    author = "Tumasyan, Armen and others",
    collaboration = "CMS",
    title = "{Observation of $\tau$ lepton pair production in ultraperipheral lead-lead collisions at $\sqrt{s_\mathrm{NN}}$ = 5.02 TeV}",
    eprint = "2206.05192",
    archivePrefix = "arXiv",
    primaryClass = "nucl-ex",
    reportNumber = "CMS-HIN-21-009, CERN-EP-2022-098",
    doi = "10.1103/PhysRevLett.131.151803",
    journal = "Phys. Rev. Lett.",
    volume = "131",
    pages = "151803",
    year = "2023"
}

@article{ATLAS:2022ryk,
    author = "Aad, Georges and others",
    collaboration = "ATLAS",
    title = "{Observation of the {\ensuremath{\gamma}}{\ensuremath{\gamma}}{\textrightarrow}{\ensuremath{\tau}}{\ensuremath{\tau}} Process in Pb+Pb Collisions and Constraints on the {\ensuremath{\tau}}-Lepton Anomalous Magnetic Moment with the ATLAS Detector}",
    eprint = "2204.13478",
    archivePrefix = "arXiv",
    primaryClass = "hep-ex",
    reportNumber = "CERN-EP-2022-079",
    doi = "10.1103/PhysRevLett.131.151802",
    journal = "Phys. Rev. Lett.",
    volume = "131",
    number = "15",
    pages = "151802",
    year = "2023"
}

@article{CMS:2024qjo,
    author = "Hayrapetyan, Aram and others",
    collaboration = "CMS",
    title = "{Observation of $\gamma\gamma\to\tau\tau$ in proton-proton collisions and limits on the anomalous electromagnetic moments of the $\tau$ lepton}",
    eprint = "2406.03975",
    archivePrefix = "arXiv",
    primaryClass = "hep-ex",
    reportNumber = "CMS-SMP-23-005, CERN-EP-2024-127",
    doi = "10.1088/1361-6633/ad6fcb",
    journal = "Rept. Prog. Phys.",
    volume = "87",
    number = "10",
    pages = "107801",
    year = "2024"
}

@article{Bierlich:2024vqo,
    author = "Bierlich, Christian and Buckley, Andy and Butterworth, Jonathan Mark and Gutschow, Christian and Lonnblad, Leif and Procter, Tomasz and Richardson, Peter and Yeh, Yoran",
    title = "{Robust independent validation of experiment and theory: Rivet version 4 release note}",
    eprint = "2404.15984",
    archivePrefix = "arXiv",
    primaryClass = "hep-ph",
    reportNumber = "MCNET-24-05",
    doi = "10.21468/SciPostPhysCodeb.36",
    journal = "SciPost Phys. Codeb.",
    volume = "36",
    pages = "1",
    year = "2024"
}

@article{Fermi:1924tc,
    author = "Fermi, E.",
    title = "{On the Theory of the impact between atoms and electrically charged particles}",
    doi = "10.1007/BF03184853",
    journal = "Z. Phys.",
    volume = "29",
    pages = "315--327",
    year = "1924"
}

@article{vonWeizsacker:1934nji,
    author = "von Weizsacker, C. F.",
    title = "{Radiation emitted in collisions of very fast electrons}",
    doi = "10.1007/BF01333110",
    journal = "Z. Phys.",
    volume = "88",
    pages = "612--625",
    year = "1934"
}

@article{Williams:1934ad,
    author = "Williams, E. J.",
    title = "{Nature of the high-energy particles of penetrating radiation and status of ionization and radiation formulae}",
    doi = "10.1103/PhysRev.45.729",
    journal = "Phys. Rev.",
    volume = "45",
    pages = "729--730",
    year = "1934"
}

@article{Budnev:1975poe,
    author = "Budnev, V. M. and Ginzburg, I. F. and Meledin, G. V. and Serbo, V. G.",
    title = "{The Two photon particle production mechanism. Physical problems. Applications. Equivalent photon approximation}",
    doi = "10.1016/0370-1573(75)90009-5",
    journal = "Phys. Rept.",
    volume = "15",
    pages = "181--281",
    year = "1975"
}

@article{Krauss:1997vr,
    author = "Krauss, F. and Greiner, M. and Soff, G.",
    title = "{Photon and gluon induced processes in relativistic heavy ion collisions}",
    doi = "10.1016/S0146-6410(97)00049-5",
    journal = "Prog. Part. Nucl. Phys.",
    volume = "39",
    pages = "503--564",
    year = "1997"
}

@article{sidi1988user,
  title={A user-friendly extrapolation method for oscillatory infinite integrals},
  author={Sidi, Avram},
  journal={Mathematics of Computation},
  volume={51},
  number={183},
  pages={249--266},
  year={1988}
}

@article{lucas1995evaluating,
  title={Evaluating infinite integrals involving Bessel functions of arbitrary order},
  author={Lucas, SK and Stone, HA},
  journal={Journal of Computational and Applied Mathematics},
  volume={64},
  number={3},
  pages={217--231},
  year={1995},
  publisher={Elsevier}
}

@article{Klein:2016yzr,
    author = "Klein, Spencer R. and Nystrand, Joakim and Seger, Janet and Gorbunov, Yuri and Butterworth, Joey",
    title = "{STARlight: A Monte Carlo simulation program for ultra-peripheral collisions of relativistic ions}",
    eprint = "1607.03838",
    archivePrefix = "arXiv",
    primaryClass = "hep-ph",
    doi = "10.1016/j.cpc.2016.10.016",
    journal = "Comput. Phys. Commun.",
    volume = "212",
    pages = "258--268",
    year = "2017"
}

@article{Harland-Lang:2020veo,
    author = "Harland-Lang, L. A. and Tasevsky, M. and Khoze, V. A. and Ryskin, M. G.",
    title = "{A new approach to modelling elastic and inelastic photon-initiated production at the LHC: SuperChic 4}",
    eprint = "2007.12704",
    archivePrefix = "arXiv",
    primaryClass = "hep-ph",
    reportNumber = "IPPP/20/33",
    doi = "10.1140/epjc/s10052-020-08455-0",
    journal = "Eur. Phys. J. C",
    volume = "80",
    number = "10",
    pages = "925",
    year = "2020"
}

@article{Sherpa:2024mfk,
    author = "Bothmann, Enrico and others",
    collaboration = "Sherpa",
    title = "{Event generation with Sherpa 3}",
    eprint = "2410.22148",
    archivePrefix = "arXiv",
    primaryClass = "hep-ph",
    reportNumber = "IPPP/24/67, LTH-1385, FERMILAB-PUB-24-0748-T, ZU-TH 51/24, MCNET-24-17, CERN-TH-2024-171",
    doi = "10.1007/JHEP12(2024)156",
    journal = "JHEP",
    volume = "12",
    pages = "156",
    year = "2024"
}

@article{Krauss:2001iv,
    author = "Krauss, F. and Kuhn, R. and Soff, G.",
    title = "{AMEGIC++ 1.0: A Matrix element generator in C++}",
    eprint = "hep-ph/0109036",
    archivePrefix = "arXiv",
    reportNumber = "CAVENDISH-HEP-01-11",
    doi = "10.1088/1126-6708/2002/02/044",
    journal = "JHEP",
    volume = "02",
    pages = "044",
    year = "2002"
}

@article{Yennie:1961ad,
    author = "Yennie, D. R. and Frautschi, Steven C. and Suura, H.",
    title = "{The infrared divergence phenomena and high-energy processes}",
    doi = "10.1016/0003-4916(61)90151-8",
    journal = "Annals Phys.",
    volume = "13",
    pages = "379--452",
    year = "1961"
}

@article{Catani:1996vz,
    author = "Catani, S. and Seymour, M. H.",
    title = "{A General algorithm for calculating jet cross-sections in NLO QCD}",
    eprint = "hep-ph/9605323",
    archivePrefix = "arXiv",
    reportNumber = "CERN-TH-96-029, CERN-TH-96-29",
    doi = "10.1016/S0550-3213(96)00589-5",
    journal = "Nucl. Phys. B",
    volume = "485",
    pages = "291--419",
    year = "1997",
    note = "[Erratum: Nucl.Phys.B 510, 503--504 (1998)]"
}

@article{Buckley:2019xhk,
    author = {Buckley, Andy and Ilten, Philip and Konstantinov, Dmitri and L{\"o}nnblad, Leif and Monk, James and Pokorski, Witold and Przedzinski, Tomasz and Verbytskyi, Andrii},
    title = "{The HepMC3 event record library for Monte Carlo event generators}",
    eprint = "1912.08005",
    archivePrefix = "arXiv",
    primaryClass = "hep-ph",
    reportNumber = "MPP-2019-258, MCNET-19-27, LU-TP 19-58",
    doi = "10.1016/j.cpc.2020.107310",
    journal = "Comput. Phys. Commun.",
    volume = "260",
    pages = "107310",
    year = "2021"
}

@article{Schonherr:2017qcj,
    author = {Sch{\"o}nherr, Marek},
    title = "{An automated subtraction of NLO EW infrared divergences}",
    eprint = "1712.07975",
    archivePrefix = "arXiv",
    primaryClass = "hep-ph",
    reportNumber = "CERN-TH-17-281, CERN-TH-2017-281, MCNET-17-24",
    doi = "10.1140/epjc/s10052-018-5600-z",
    journal = "Eur. Phys. J. C",
    volume = "78",
    number = "2",
    pages = "119",
    year = "2018"
}

@article{Schonherr:2008av,
    author = "Schonherr, Marek and Krauss, Frank",
    title = "{Soft Photon Radiation in Particle Decays in SHERPA}",
    eprint = "0810.5071",
    archivePrefix = "arXiv",
    primaryClass = "hep-ph",
    reportNumber = "DCPT-07-96, IPPP-07-48, MCNET-08-13",
    doi = "10.1088/1126-6708/2008/12/018",
    journal = "JHEP",
    volume = "12",
    pages = "018",
    year = "2008"
}

@article{Gleisberg:2008fv,
    author = "Gleisberg, Tanju and Hoeche, Stefan",
    title = "{Comix, a new matrix element generator}",
    eprint = "0808.3674",
    archivePrefix = "arXiv",
    primaryClass = "hep-ph",
    reportNumber = "SLAC-PUB-13232, IPPP-08-31, DCPT-08-62, MCNET-08-08",
    doi = "10.1088/1126-6708/2008/12/039",
    journal = "JHEP",
    volume = "12",
    pages = "039",
    year = "2008"
}

@article{Gleisberg:2007md,
    author = "Gleisberg, Tanju and Krauss, Frank",
    title = "{Automating dipole subtraction for QCD NLO calculations}",
    eprint = "0709.2881",
    archivePrefix = "arXiv",
    primaryClass = "hep-ph",
    reportNumber = "DCPT-07-88, IPPP-07-44",
    doi = "10.1140/epjc/s10052-007-0495-0",
    journal = "Eur. Phys. J. C",
    volume = "53",
    pages = "501--523",
    year = "2008"
}

@article{Bertulani:2005ru,
    author = "Bertulani, Carlos A. and Klein, Spencer R. and Nystrand, Joakim",
    title = "{Physics of ultra-peripheral nuclear collisions}",
    eprint = "nucl-ex/0502005",
    archivePrefix = "arXiv",
    doi = "10.1146/annurev.nucl.55.090704.151526",
    journal = "Ann. Rev. Nucl. Part. Sci.",
    volume = "55",
    pages = "271--310",
    year = "2005"
}

@article{Dittmaier:2025ikh,
    author = "Dittmaier, Stefan and Engel, Tim and Ariza, Jose Luis Hernando and Pellen, Mathieu",
    title = "{Electroweak corrections to {\ensuremath{\tau}}$^{+}${\ensuremath{\tau}}$^{-}$ production in ultraperipheral heavy-ion collisions at the LHC}",
    eprint = "2504.11391",
    archivePrefix = "arXiv",
    primaryClass = "hep-ph",
    reportNumber = "FR-PHENO-2025-004",
    doi = "10.1007/JHEP08(2025)051",
    journal = "JHEP",
    volume = "08",
    pages = "051",
    year = "2025"
}

@article{Shao:2024dmk,
    author = "Shao, Hua-Sheng and d'Enterria, David",
    title = "{Dimuon and ditau production in photon-photon collisions at next-to-leading order in QED}",
    eprint = "2407.13610",
    archivePrefix = "arXiv",
    primaryClass = "hep-ph",
    doi = "10.1007/JHEP02(2025)023",
    journal = "JHEP",
    volume = "02",
    pages = "023",
    year = "2025"
}

@article{Harland-Lang:2021ysd,
    author = "Harland-Lang, L. A. and Khoze, V. A. and Ryskin, M. G.",
    title = "{Elastic photon-initiated production at the LHC: the role of hadron-hadron interactions}",
    eprint = "2104.13392",
    archivePrefix = "arXiv",
    primaryClass = "hep-ph",
    reportNumber = "IPPP/20/97",
    doi = "10.21468/SciPostPhys.11.3.064",
    journal = "SciPost Phys.",
    volume = "11",
    pages = "064",
    year = "2021"
}

@article{Burmasov:2021phy,
    author = "Burmasov, Nazar and Kryshen, Evgeny and Buehler, Paul and Lavicka, Roman",
    title = "{Upcgen: A Monte Carlo simulation program for dilepton pair production in ultra-peripheral collisions of heavy ions}",
    eprint = "2111.11383",
    archivePrefix = "arXiv",
    primaryClass = "hep-ph",
    doi = "10.1016/j.cpc.2022.108388",
    journal = "Comput. Phys. Commun.",
    volume = "277",
    pages = "108388",
    year = "2022"
}

@article{Bailey:2022wqy,
    author = "Bailey, S. and Harland-Lang, L. A.",
    title = "{Modeling W+W- production with rapidity gaps at the LHC}",
    eprint = "2201.08403",
    archivePrefix = "arXiv",
    primaryClass = "hep-ph",
    doi = "10.1103/PhysRevD.105.093010",
    journal = "Phys. Rev. D",
    volume = "105",
    number = "9",
    pages = "093010",
    year = "2022"
}

@article{Harland-Lang:2024zpn,
    author = "Harland-Lang, L. A.",
    title = "{Higher precision constraints on the tau $g-2$ in LHC photon-initiated production: a full account of hadron dissociation and soft survival effects}",
    eprint = "2410.10978",
    archivePrefix = "arXiv",
    primaryClass = "hep-ph",
    doi = "10.1140/epjc/s10052-024-13685-7",
    journal = "Eur. Phys. J. C",
    volume = "84",
    number = "12",
    pages = "1332",
    year = "2024"
}

@article{ATLAS:2016lse,
    author = "Aaboud, Morad and others",
    collaboration = "ATLAS",
    title = "{Measurement of exclusive $\gamma\gamma\rightarrow W^+W^-$ production and search for exclusive Higgs boson production in $pp$ collisions at $\sqrt{s} = 8$ TeV using the ATLAS detector}",
    eprint = "1607.03745",
    archivePrefix = "arXiv",
    primaryClass = "hep-ex",
    reportNumber = "CERN-EP-2016-123",
    doi = "10.1103/PhysRevD.94.032011",
    journal = "Phys. Rev. D",
    volume = "94",
    number = "3",
    pages = "032011",
    year = "2016"
}

@article{ATLAS:2026wrz,
    author = "Aad, Georges and others",
    collaboration = "ATLAS",
    title = "{Differential measurements of {\ensuremath{\gamma}}{\ensuremath{\gamma}}{\textrightarrow}{\ensuremath{\tau}}{\ensuremath{\tau}} and constraints on $\tau$-lepton electromagnetic moments in Pb+Pb collisions at $\sqrt{s_{_\text{NN}}} = 5.02$ TeV with ATLAS}",
    eprint = "2605.04584",
    archivePrefix = "arXiv",
    primaryClass = "nucl-ex",
    reportNumber = "CERN-EP-2026-135",
    month = "5",
    year = "2026"
}

@article{Guzey:2018dlm,
    author = "Guzey, V. and Klasen, M.",
    title = "{Inclusive dijet photoproduction in ultraperipheral heavy ion collisions at the CERN Large Hadron Collider in next-to-leading order QCD}",
    eprint = "1811.10236",
    archivePrefix = "arXiv",
    primaryClass = "hep-ph",
    reportNumber = "MS-TP-18-30",
    doi = "10.1103/PhysRevC.99.065202",
    journal = "Phys. Rev. C",
    volume = "99",
    number = "6",
    pages = "065202",
    year = "2019"
}

@article{Jiang:2024dhf,
    author = "Jiang, Jun and Lu, Peng-Cheng and Si, Zong-Guo and Zhang, Han and Zhang, Xin-Yi",
    title = "{NLO EW corrections to tau pair production via photon fusion in Pb-Pb ultraperipheral collisions}",
    eprint = "2410.21963",
    archivePrefix = "arXiv",
    primaryClass = "hep-ph",
    doi = "10.1103/PhysRevD.111.036023",
    journal = "Phys. Rev. D",
    volume = "111",
    number = "3",
    pages = "036023",
    year = "2025"
}

@article{Bertulani:1987tz,
    author = "Bertulani, Carlos A. and Baur, Gerhard",
    title = "{Electromagnetic Processes in Relativistic Heavy Ion Collisions}",
    reportNumber = "JUL-2163",
    doi = "10.1016/0370-1573(88)90142-1",
    journal = "Phys. Rept.",
    volume = "163",
    pages = "299",
    year = "1988"
}

@article{Sjostrand:2006za,
    author = "Sjostrand, Torbjorn and Mrenna, Stephen and Skands, Peter Z.",
    title = "{PYTHIA 6.4 Physics and Manual}",
    eprint = "hep-ph/0603175",
    archivePrefix = "arXiv",
    reportNumber = "FERMILAB-PUB-06-052-CD-T, LU-TP-06-13",
    doi = "10.1088/1126-6708/2006/05/026",
    journal = "JHEP",
    volume = "05",
    pages = "026",
    year = "2006"
}

@article{Bierlich:2022pfr,
    author = "Bierlich, Christian and others",
    title = "{A comprehensive guide to the physics and usage of PYTHIA 8.3}",
    eprint = "2203.11601",
    archivePrefix = "arXiv",
    primaryClass = "hep-ph",
    reportNumber = "LU-TP 22-16, MCNET-22-04, FERMILAB-PUB-22-227-SCD",
    doi = "10.21468/SciPostPhysCodeb.8",
    journal = "SciPost Phys. Codeb.",
    volume = "2022",
    pages = "8",
    year = "2022"
}

@article{Bellm:2025pcw,
    author = "Bellm, J. and others",
    title = "{The Physics of Herwig 7}",
    eprint = "2512.16645",
    archivePrefix = "arXiv",
    primaryClass = "hep-ph",
    reportNumber = "CERN-TH-2025-252, IPPP-25-57, HERWIG-2025-01, KA-TP-36-2025, MCNET-25-31",
    month = "12",
    year = "2025"
}

@article{Dyndal:2026uvm,
    author = "Dyndal, M. and Harland-Lang, L. A.",
    title = "{A First Account of the Impact of Ion Electromagnetic Dissociation on Event Exclusivity in Ultraperipheral LHC Collisions}",
    eprint = "2604.19879",
    archivePrefix = "arXiv",
    primaryClass = "hep-ph",
    month = "4",
    year = "2026"
}

@unpublished{photonuclear-jets,
  author          = "Bothmann, E. and Meinzinger, P.",
  title           = "{Photo-nuclear jets at NLO+PS accuracy}",
  note            = "in preparation",
  year            = {2026}
}

@article{Vidovic:1992ik,
    author = "Vidovic, M. and Greiner, M. and Best, C. and Soff, G.",
    title = "{Impact parameter dependence of the electromagnetic particle production in ultrarelativistic heavy ion collisions}",
    reportNumber = "GSI-92-52",
    doi = "10.1103/PhysRevC.47.2308",
    journal = "Phys. Rev. C",
    volume = "47",
    pages = "2308--2319",
    year = "1993"
}

@article{Vidovic:1993cf,
    author = "Vidovic, Mario and Greiner, Martin and Soff, Gerhard",
    title = "{Electromagnetic dissociation of Pb nuclei in peripheral ultrarelativistic heavy ion collisions}",
    reportNumber = "GSI-93-34",
    doi = "10.1103/PhysRevC.48.2011",
    journal = "Phys. Rev. C",
    volume = "48",
    pages = "2011--2015",
    year = "1993"
}

@article{Shao:2022cly,
    author = "Shao, Hua-Sheng and d'Enterria, David",
    title = "{gamma-UPC: automated generation of exclusive photon-photon processes in ultraperipheral proton and nuclear collisions with varying form factors}",
    eprint = "2207.03012",
    archivePrefix = "arXiv",
    primaryClass = "hep-ph",
    doi = "10.1007/JHEP09(2022)248",
    journal = "JHEP",
    volume = "09",
    pages = "248",
    year = "2022"
}

@article{Dyndal:2014yea,
    author = "Dyndal, Mateusz and Schoeffel, Laurent",
    title = "{The role of finite-size effects on the spectrum of equivalent photons in proton{\textendash}proton collisions at the LHC}",
    eprint = "1410.2983",
    archivePrefix = "arXiv",
    primaryClass = "hep-ph",
    doi = "10.1016/j.physletb.2014.12.019",
    journal = "Phys. Lett. B",
    volume = "741",
    pages = "66--70",
    year = "2015"
}

@article{ATLAS:2017sfe,
    author = "Aaboud, Morad and others",
    collaboration = "ATLAS",
    title = "{Measurement of the exclusive $\gamma \gamma \rightarrow \mu^+ \mu^-$ process in proton-proton collisions at $\sqrt{s}=13$ TeV with the ATLAS detector}",
    eprint = "1708.04053",
    archivePrefix = "arXiv",
    primaryClass = "hep-ex",
    reportNumber = "CERN-EP-2017-151",
    doi = "10.1016/j.physletb.2017.12.043",
    journal = "Phys. Lett. B",
    volume = "777",
    pages = "303--323",
    year = "2018"
}

@article{ATLAS:2020epq,
    author = "Aad, Georges and others",
    collaboration = "ATLAS",
    title = "{Exclusive dimuon production in ultraperipheral Pb+Pb collisions at $\sqrt{s_{\mathrm{NN}}} = 5.02$ TeV with ATLAS}",
    eprint = "2011.12211",
    archivePrefix = "arXiv",
    primaryClass = "nucl-ex",
    reportNumber = "CERN-EP-2020-138",
    doi = "10.1103/PhysRevC.104.024906",
    journal = "Phys. Rev. C",
    volume = "104",
    pages = "024906",
    year = "2021"
}

@article{Klein:1999qj,
    author = "Klein, Spencer and Nystrand, Joakim",
    title = "{Exclusive vector meson production in relativistic heavy ion collisions}",
    eprint = "hep-ph/9902259",
    archivePrefix = "arXiv",
    reportNumber = "LBNL-42768, LBL-42768",
    doi = "10.1103/PhysRevC.60.014903",
    journal = "Phys. Rev. C",
    volume = "60",
    pages = "014903",
    year = "1999"
}

@article{ParticleDataGroup:2024cfk,
    author = "Navas, S. and others",
    collaboration = "Particle Data Group",
    title = "{Review of particle physics}",
    doi = "10.1103/PhysRevD.110.030001",
    journal = "Phys. Rev. D",
    volume = "110",
    number = "3",
    pages = "030001",
    year = "2024"
}

@article{DeJager:1974liz,
    author = "De Jager, C. W. and De Vries, H. and De Vries, C.",
    title = "{Nuclear charge and magnetization density distribution parameters from elastic electron scattering}",
    doi = "10.1016/S0092-640X(74)80002-1",
    journal = "Atom. Data Nucl. Data Tabl.",
    volume = "14",
    pages = "479--508",
    year = "1974",
    note = "[Erratum: Atom.Data Nucl.Data Tabl. 16, 580--580 (1975)]"
}

@article{DeVries:1987atn,
    author = "De Vries, H. and De Jager, C. W. and De Vries, C.",
    title = "{Nuclear charge and magnetization density distribution parameters from elastic electron scattering}",
    doi = "10.1016/0092-640X(87)90013-1",
    journal = "Atom. Data Nucl. Data Tabl.",
    volume = "36",
    pages = "495--536",
    year = "1987"
}

@article{Eskola:2024fhf,
    author = "Eskola, Kari J. and Guzey, Vadim and Helenius, Ilkka and Paakkinen, Petja and Paukkunen, Hannu",
    title = "{Spatial resolution of dijet photoproduction in near-encounter ultraperipheral nuclear collisions}",
    eprint = "2404.09731",
    archivePrefix = "arXiv",
    primaryClass = "hep-ph",
    doi = "10.1103/PhysRevC.110.054906",
    journal = "Phys. Rev. C",
    volume = "110",
    number = "5",
    pages = "054906",
    year = "2024"
}

@article{Dudek:1981zz,
    author = "Dudek, J. and Szymanski, Z. and Werner, Tomasz R.",
    title = "{Woods-Saxon potential parameters optimized to the high spin spectra in the lead region}",
    doi = "10.1103/PhysRevC.23.920",
    journal = "Phys. Rev. C",
    volume = "23",
    pages = "920--925",
    year = "1981"
}
\end{document}